\documentclass[pdflatex,sn-mathphys-num]{sn-jnl}

\usepackage{graphicx}%
\usepackage{multirow}%
\usepackage{amsmath,amssymb,amsfonts}%
\usepackage{amsthm}%
\usepackage{mathrsfs}%
\usepackage[title]{appendix}%
\usepackage{xcolor}%
\usepackage{textcomp}%
\usepackage{manyfoot}%
\usepackage{booktabs}%
\usepackage{algorithm}%
\usepackage{algorithmicx}%
\usepackage{algpseudocode}%
\usepackage{listings}%
\usepackage{float}

\begin{document}
\title[Marketing time series modeling]{Multifractal wavelet dynamic mode decomposition modeling for marketing time series}
\author[1]{\fnm{Mohamed Elshazli A.} \sur{Zidan}}\email{sho.shoza@gmail.com}
\author*[1,2,3]{\fnm{Anouar} \sur{Ben Mabrouk}}\email{anouar.benmabrouk@fsm.rnu.tn}
\author[4,5]{\fnm{Nidhal} \sur{Ben Abdallah}}\email{nidhal.benabdallah@fsegma.rnu.tn}
\author[6]{\fnm{Tawfeeq M.} \sur{Alanazi}}\email{t.aalenazi@ut.edu.sa}
\equalcont{These authors contributed equally to this work.}
\affil*[1]{\orgdiv{Department of Mathematics, Faculty of Science}, \orgname{University of Tabuk}, \orgaddress{\street{King Faisal road}, \city{Tabuk}, \postcode{47512}, \country{Saudi Arabia}}}
\affil[2]{\orgdiv{Department of Mathematics, Higher Institute of Applied Mathematics and Computer Science}, \orgname{University of Kairouan}, \orgaddress{\street{Street of Assad Ibn Alfourat}, \city{Kairouan}, \postcode{3100}, \country{Tunisia}}}
\affil[3]{\orgdiv{Laboratory of Algebra, Number Theory and Nonlinear Analysis, Department of Mathematics, Faculty of Sciences}, \orgname{University of Monastir}, \orgaddress{\street{Avenue of the Environment}, \city{Monastir}, \postcode{5000}, \country{Tunisia}}}
\affil[4]{\orgdiv{Department of Quantitative Methods, Faculty of Economic Sciences and Management}, \orgname{University of Monastir}, \orgaddress{\street{Sidi Messaoud}, \city{Hiboun}, \postcode{5111}, \state{Mahdia}, \country{Tunisia}}}
\affil[5]{\orgdiv{Laboratoire THEMA de l'ESCT Tunis}, \orgname{Campus Universitaire de Manouba}, \orgaddress{\street{Université de Carthage}, \city{Mannouba}, \postcode{2010}, \state{Mannouba}, \country{Tunisia}}}
\affil[6]{\orgdiv{Department of Marketing, Faculty of Business Administration}, \orgname{University of Tabuk}, \orgaddress{\street{King Faisal road}, \city{Tabuk}, \postcode{47512}, \country{Saudi Arabia}}}
\abstract{Marketing is the way we ensure our sales are the best in the market, our prices the most accessible, and our clients satisfied, thus ensuring our brand has the widest distribution. This requires sophisticated and advanced understanding of the whole related network. Indeed, marketing data may exist in different forms such as qualitative and quantitative data. However, in the literature, it is easily noted that large bibliographies may be collected about qualitative studies, while only a few studies adopt a quantitative point of view. This is a major drawback that results in marketing science still focusing on design, although the market is strongly dependent on quantities such as money and time. Indeed, marketing data may form time series such as brand sales in specified periods, brand-related prices over specified periods, market shares, etc. The purpose of the present work is to investigate some marketing models based on time series for various brands. This paper aims to combine the dynamic mode decomposition and wavelet decomposition to study marketing series due to both prices, and volume sales in order to explore the effect of the time scale on the persistence of brand sales in the market and on the forecasting of such persistence, according to the characteristics of the brand and the related market competition or competitors. Our study is based on a sample of Saudi brands during the period 22 November 2017 to 30 December 2021.}
\maketitle
\section{Introduction}
The present decade has known many severe moments/crises, social, political, and economic due to many dramatic movements such as the so-called Arab Spring, the financial crisis, COVID-19 pandemic, and lastly the war between Russia and Ukraine, which directly affected the energy market and thus the economy's stability in quasi all the world. These facts need sophisticated tools to understand the situation of the market as a first player and main factor in the country's stability, as well as the correct policy and government. 

Understanding the market is based on a good understanding and/or analysis of the data it characterized or extracted. Generally, a good data analysis needs scientific computation tools and high-fidelity measurements, especially in the stock market where the data may hide dynamics, anomalies, singularities, extreme values, ..., etc. This induces immediate capabilities for prediction, state estimation, and also good control of the whole system.

In this paper, our focus will be on developing a mathematical technique geared toward analyzing financial time series issued from marketing brands by combining the so-called dynamic mode decomposition (DMD) and the wavelet decomposition (WD). 

The aim of this work is to investigate a wavelet-based mathematical model for brand sales and prices. Wavelet theory has gained great interest in the fields of economics and finance since the last decades of the last century. Recall that almost 
 all financial indices usually show volatile behavior as well as high fluctuations. As a result, their investigation using classical methods remains insufficient. Wavelets have indeed been shown to be useful in such~investigations.

Marketing, in its simplest sense, includes the institutions, the activities, and also the processes applied to deliver, create, and exchange goods, and to communicate offerings and needs to customers, partners, and clients, or more broadly to society \cite{Armstrong}. While marketing consists of various activities, the principles of marketing rely on four elements, i.e., product, price, promotion, and place (distribution). Each one of these elements depends on qualitative and quantitative data. Quantitative data show the power of marketing in collecting, analyzing, optimizing, and executing, based on relevant numerical data. Product choices and portfolios, price hierarchies, numbers of outlets and stores, and, importantly, promotion budgets, especially for branding, brand management, and brand equity are also considered. Brands and the activity of branding are well represented in the literature, 
 and brand equity refers to the value added to products \cite{Jones}. Firms can benefit from a strong brand in various ways: it improves the product value and provides a chance to extend the brand to new products, expand into new markets, and obtain a high level of customer loyalty and tolerance. A well-known brand acquires a price value \cite{Farquhar}. Thus, brand sales- and equity-related data include things such as choice data, the brand design, the brand idea, time of purchases, the number of stores, the choice of price, and the competitors in the market. The state of the market itself affects the persistence and the success of the brand investment, strongly and simultaneously. Without sophisticated mathematical models that take into account all the factors and that permit a good understanding of the time and space evolution of the brand distribution and its evolution in the market, the investment in the brand may fail. 

In marketing, a mathematical model may include many factors such as the number of sales in specified periods such as weeks, months, or years, market shares per specified period, price evolution, the history of the market and its situation, competing brands, etc. The main characteristic of a mathematical model is its ability to deal with many factors simultaneously, according to time and space. It gathers the data into a time series form, permitting ordering over time. This has a very important consequence, as it yields information about the future state of the brand. The model uses a time series, as the best descriptor to extract, deduce, and/or accurately describe the properties of the empirical data.

The DMD method is a matrix decomposition technique used to discover the singular values in a time series. However, the drawback of this method is related to the low-rank structures extracted from it. These are associated with temporal features as well as correlated spatial activity. To understand the importance of DMD, researchers used the well-known principal component analysis (PCA) and also the Fourier transform (See \cite{KutzDMDBook}). 

The WD is a mathematical tool developed theoretically since the 1980's era due to petroleum extraction, where the Fourier transform failed to explain well the extracted signal. The WD is characterized by its capability to describe, model, and analyze many data sets issued from different fields such as signals, images, videos, bio-signals, statistics, finance, ... etc. It is also proved to be a good approximation and estimator in theoretical functional and statistical analysis. The principle of WD is based on a representation of the analyzed object into a series of modes due to special functions called wavelets issued from one source known as the wavelet mother/father by dilation and translation permitting to reach the singularities around any point in the whole time or space domain (See \cite{Arfaoui1,Arfaoui2,Gencay,Hardle,Percival}).

The next section concerns a brief literature review of existing studies on marketing time series models. Section 3 is devoted to the presentation of the dynamic mode decomposition and the wavelet decomposition methods for the marketing time series processing and modeling. Section 4 concerns the empirical results, together with discussions and interpretations. Section 5 concludes the paper.
\section{Literature Review}\label{sec2}
Generally, marketing researchers and/or modelers search for models taking into account one or more variables solely as functions of time. Next, the model is analyzed regarding its efficiency for interpolating (approximating) and extrapolating (predicting) these variables in time. 

Compared to other fields such as financial and economic applications, the use of mathematical models, especially wavelet-based models, is still not well developed in marketing. In \cite{Dekimpe2000}, for example, the authors mention the difficulty of time series  training for traditional research models, the lack of software adapted to marketing modeling, the lack of, and sometimes the poor quality of, time-series data, and the lack of a substantive area for marketing, allowing time-series-based models to be readily adopted. 

The problem of time scales in marketing has been raised in several studies. In \cite{Michis2009a}, it is mentioned that marketing phenomena depend on frequencies in time relative to decisions and interactions in the market. In classical investigations, the most commonly used time intervals are weekly, monthly, quarterly, and annual cycles. Each period has its special and/or specific characteristics. Weekly cycles, for example, are for reductions in prices, and quarterly periods are generally related to regular price adjustments and~competition. 

Moreover, many phenomena such as pandemics (COVID-19), wars, climate change, financial crises, and socio-political movements may induce severe perturbations in the usual considerations, especially time periods, leading to diversification in the use of time~scales.

This diversification in how to choose the time scale, according to what factors, what characteristics, and what aim, has led researchers in the marketing field to think about including the time scale strongly in the models. In \cite{Leeflangetal2000}, the authors applied certain types of short-, medium-, and long-term horizons to investigate the effect of the time scale. However, their choice was not adapted to the wavelet method.

In marketing, according to \cite{Michis2009a,Michis2009b}, the time scale is mostly applied at discrete, equally spaced intervals, with weekly periods dominating, although this hypothesis has been criticized. In \cite{Michis2009b}, the author applied a wavelet-based method for forecasting brand sales. The problem of multicollinearity has been investigated, resulting in correlated vectors of coefficients which have been applied to provide the most accurate forecasting and the best dimension reduction. In \cite{Leone1995}, for example, it is mentioned that marketing models still present aggregations, even using weekly intervals. This allows us to conclude that more adequate tools for time/frequency scales should be applied. 

Many methods have been used in this context, such as Fourier analysis and spectral analysis. However, these tools have raised many problems such as the non-stationary behavior of the marketing data, the sizes of samples, etc. (\cite{Bronnenberg,Dekimpe95b,Deleersnyder,Kaiser1999,Kaiser2001,Pauwels2004}).

Wavelet analysis was introduced into marketing models to overcome limitations such as the non-localization aspect of spectral analysis and its concentration in the frequency domain. This is achieved by frequency decomposition of the statistical time series into components that are well localized in time/frequency.

Michis mentioned a very important characteristic of wavelet theory in \cite{Michis2007,Michis2009a,Michis2009b}. It permits investigation of the causal links between time-series cycles in marketing according to time scale. The modelers differentiate the marketing-driver influences in long-duration cycles of sales. Wavelets are also good tools regarding estimation and prediction accuracy and the handling of non-stationary time series. The application of wavelets in marketing resides in covering/uncovering the frequency activity characteristics of marketing models. Wavelet crystals generated from the inverse of the wavelet transform are applied to localize the variation in economic variables over different horizons. Recall that brand sales forecasting is important for investors and consumers but also for modelers, due to its strong relationships with financial resources allocation and budget planning. Therefore, a successful sales forecast will yield a good guide to planning sufficient production and good product distribution, according to both time and demand. Furthermore, appropriate marketing activities will be well organized, resulting in good performance and placement in the market \cite{Michis2007,Michis2009a,Michis2009b}.
\section{The DMD and WD tools for time series}
The WD and the DMD act by decomposing complex data into simpler modes based on spatio-temporal coherent structures to be able to provide insights from such data. This permits for example to best predict the behavior of the data and the real phenomenon behind it. 

Let $X_i=X(t_i)$, $t=t_1,t_2,...,t_N$ be a time series due to a dynamical system, for example, the DMD method seeks the best model 
\begin{equation}\label{eq1}
X_{k+1}=F(X_i,i\leq k)
\end{equation}
that regresses the best the data locally by minimizing an $L^2$-error $\|X_{k+1}-F(X_i,i\leq k)\|_2$ over the time interval $k=1,2,3,...,N-1$. 

In the case of a discrete linear dynamical system $X_{k+1}=AX_k$, $A$ being a fixed matrix, the solution may be expressed by means of the eigenvalues and eigenvectors of the matrix $A$, and the initial solution as
\begin{equation}\label{eq2}
X_k=\displaystyle\sum_i\Phi_i\lambda_i^kX_{0,i},
\end{equation}
where the $\Phi_i$ are the eigenvectors, the $\lambda_i$ are the eigenvalues, and $X_{0,i}$ are the coordinates of the initial solution $X_0$.

Notice from equation (\ref{eq2}) that the DMD uses in some way a functional basis to estimate the solution. At short-time horizons, we may approximate the eigenvalues and the eigenvectors by setting $\omega_k=\ln\lambda_k/\Delta t$, to get an approximation of the solution 
\begin{equation}\label{eq3}
X(t)=\displaystyle\sum_i\Phi_i\exp(\omega_it)X_{0,i},
\end{equation}
which confirms the use of a suitable function basis (or modes), such as the Fourier one in (\ref{eq3}). Notice that such a basis permits to localize the dynamic behavior of the time series $X(t)$. 

Wavelet analysis or alternatively MRA starts with a fixed function $\psi\in L^2(\mathbb{R})$ (square integrable, or finite variance in the discrete case such as statistical series) known as the mother wavelet which generates next daughters wavelets 
\begin{equation}\label{psisp}
\psi_{s,p}(t)=\displaystyle\frac{1}{\sqrt{s}}\psi\left(\displaystyle\frac{t-p}{s}\right),    
\end{equation}
where the parameter $s>0$ is known as the scale or the scaling parameter or frequency, and $p$ stands for the position or translation. 

In wavelet analysis and/or processing, the mother wavelet $\psi$ has beneficial properties that serve for processing different cases, theoretical or practical, to yield a series representation of the data, such as the admissibility condition. 

To analyze a time (statistical) series $X(t)$, we need to evaluate its wavelet transform (called also detail coefficient) at the scale $s$ and the position $p$ through a discrete suitable grid such as the dyadic $s=2^j$, and $p=k2^j$, $j,k\in\mathbb{Z}$, by setting a discrete convolution
\begin{equation}\label{discretewavelettransformofX(t)}
DWT_{j,k}=\displaystyle\sum_{i}X(i)\psi_{j,k}(i),
\end{equation}
where we denote 
\begin{equation}\label{psijk}
\psi_{j,k}(t)=2^{-j/2}\psi(2^jt-k).
\end{equation}
This permits to reconstruct the series $X(t)$ representing the data by means of an (orthogonal) decomposition as
\begin{equation}\label{waveletseriesdecompositionofX(t)}
X(t)=\displaystyle\sum_{j\in\mathbb{Z}}\displaystyle\sum_{k\in\mathbb{Z}}DWT_{j,k}\psi_{j,k}(t).
\end{equation}
We thus split up the series $x(t)$ into a collection of subseries by translating the wavelet with the parameter $k$ over the entire time domain. Moreover, by changing the scale parameter $j$, we obtain the processing of the series $X(t)$ at different frequency bands, called also resolutions due to the scaling of the mother wavelet $\psi$ with the parameter $j$. The combination of translation and scaling allows for the processing of the signals at different times and frequencies. We thus get the so-called multiresolution analysis. For each scale (or frequency) $j$, the component
\begin{equation}\label{detailcomponentofX(t)}
DX_j=\displaystyle\sum_{k\in\mathbb{Z}}DWT_{j,k}\psi_{j,k}
\end{equation}
belongs to the special space $W_j=spann(\psi_{j,k},\,k\in\mathbb{Z})$ known in wavelet analysis as the detail space at the resolution, or the level $j$. It therefore reflects the details hidden in the data at this level. We will see on the practical cases later that effectively this component represents the dynamics, and/or the hidden behavior of the data, and that this representation differs from level to level like a zooming process. More precisely, we have an orthogonal decomposition
\begin{equation}\label{wjsommedirecte}
L^2(\mathbb{R})=\displaystyle\bigoplus^\perp_{j\in\mathbb{Z}}W_j.
\end{equation}
The multiresolution analysis consists of superposing the detail subspaces of many levels to get a whole subspace $V_j$ gathering all the details $l\leq j$ resulting thus in a global shape for the series. Denote 
\begin{equation}\label{Vj+1sommedirectedevjetwj}
V_j=\displaystyle\bigoplus^\perp_{l\leq j}W_l,
\end{equation}
we get here a sequence of nested subspaces $V_j$ of $L^2(\mathbb{R})$ which satisfies many beneficial properties such as the nesting, direct sum, zooming (\cite{Arfaoui1,Arfaoui2}). The resolution space at the level $j$ denoted by $V_j$ may be also generated analogously to $W_j$ by means of a fixed function $\Phi$ called the father wavelet or also the scaling function, for which we have $V_j=spann(\Phi_{j,k},\,k\in\mathbb{Z})$, where the $\Phi_{j,k}$ are defined from $\Phi$ by the same way as the $\psi_{j,k}$ from $\psi$ in (\ref{psijk}) yielding the approximation coefficients of the series $X(t)$ expressed as
\begin{equation}\label{approximationcoefficientsofX(t)}
A_{j,k}=\displaystyle\sum_{i}X(i)\Phi_{j,k}(i).
\end{equation}
Applying the concept of MRA, the decomposition of the data $X(t)$ into the wavelet series (\ref{waveletseriesdecompositionofX(t)}) will be splitted into parts relatively to the spaces $V_j$ as 
\begin{equation}\label{decomposition-2}
X(t) = \displaystyle\sum_{k}A_{j,k}\Phi_{j,k}(t)+\displaystyle\sum_{l\geq j+1}DX_l(t).
\end{equation}
The first part denote
\begin{equation}\label{approximationcomponentofX(t)}
AX_j(t) = \displaystyle\sum_{k}A_{j,k}\Phi_{j,k}(t)
\end{equation}
is known as the projection of the data $X(t)$ on the approximation space $V_j$, and it reflects the global shape of the data at the level $j$. In econo-financial contexts, it represents the trend of the data at the level $J$. As the parameter $j$ gets upper, this component approaches more to the observed data.

The next step to investigate the type of volatility of the time series is to test its multifractality via the multifractal wavelet spectrum. Recall that for a time series $X(t)$, this spectrum is defined as follows. Consider the $j$-level wavelet-based statistics defined for $p\in\mathbb{R}$ by
$$
\mathcal{S}_j(p)=\displaystyle\sum_{k}|d_{j,k}(X)|^p,
$$
where $ d_{j,k}(X)$ is the wavelet detail coefficient at the level $j$ and the position $k$ of the series $X(t)$. This leads to the so-called Besov exponent of the series $X(t)$ defined by
$$
b(p)=1-\displaystyle\limsup_{j\rightarrow+\infty}\displaystyle\frac{\log(\mathcal{Z}_j(p))}{j\log(2)}.
$$
The spectrum of singularities is 
$$
d(\alpha)=\displaystyle\inf_p(\alpha\,p-b(p)+1).
$$
Whenever this spectrum is a concave function, we conclude that the data under investigation is effectively multifractal or irregular volatile. 
\section{Data and results}
To show the utility of involving wavelet time scales in the mathematical model, we applied it to a typical case consisting of six brands from the set of top brands in Saudi Arabia. The choice of Saudi Arabia was justified for many reasons. Saudi Arabia is one of the biggest economies, and it is one of the main players in petroleum activities as a result of its influence on worldwide petroleum prices, and thus on economies. It is a member of the G7 group and OPEC. Saudi Arabia has also recently implemented the so-called Vision 2030, which combines many projects and activities such as marketing and which has a direct impact on both national and foreign markets, particularly via the NEOM project. The period of study may also be a strong factor affecting the results and their interpretation. Indeed, the period was related to many changes such as the GCC near-embargo against Qatar, COVID-19, the Yemen war, and all the other socio-political changes in the Arab world. The geographical position is also important, as well as the fact that the main holy cities for Muslims all over the world participate strongly in the distribution of Saudi products. In addition, Saudi Arabia and the whole GCC region is the largest workers' community in the world. With regard to the economy and marketing, we may recall that among the most important plans and aims for the Vision 2030 program are the adoption of diversity in income sources and the need to reduce the total dependence on oil, thus encouraging industrial, commercial, agricultural, and other activities. This means that the top national brands are of interest.

The brands to be considered here are given in Table \ref{TableBrands}. The chosen Saudi brands are in turn justified for many reasons, mainly the fact that these are the brands with the most worldwide distributions, and they are known in many other countries, having extended their local origins to form a worldwide reputation, consequently constituting a success for the kingdom's industry. Recall that the Kingdom of Saudi Arabia is generally known as a consumer community, which relies on foreign imported consumer products in exchange for petroleum exports. Encouraging national industry and national labor generally is one of the major goals of the Vision 2030 program. 
\begin{table}[ht]
\caption{Some top Saudi brands.}
\label{TableBrands}
\begin{tabular}{lll}
\hline
{Nomination}&{Brand}&{Description (Sector)}\\ 
$B_1$ & Jarir bookstore & Books and electronics\\
$B_2$ & Almarai & Dairy and poultry\\
$B_3$ & STC & Telecommunications\\
$B_4$ & Al Abdullatif & Household durables\\
$B_5$ & EIC & Electrical industries company\\
$B_6$ & Al Aseel & Textiles, apparel and luxury goods\\
\hline
\end{tabular}
\end{table}

Jarir Bookstore was founded in July 1974 by Abdulrahman Nasser Al-Agil. It is one of the largest retailers for books and electronics in the Kingdom of Saudi Arabia, and has now expanded to many other countries, especially those in the GCC, such as Kuwait, Qatar, and UAE. It is also one of the major components of the Saudi TADAWUL index.  

The Almarai company was originally constituted as a partnership between Prince Sultan bin Mohammed bin Saud Al Kabeer and the Irishmen Paddy McGuckian and Alastair McGuckian. Now, it is one of the biggest dairy companies in KSA and also in the whole Middle East region. It was founded 40 years ago and now specializes in dairy, poultry, juices, bakery goods, yogurt, and infant products. 

The Saudi Telecommunications Company, abbreviated STC, started 19 years ago. It basically offers telecommunications services and products. It has also now expanded into other GCC countries and to countries in other continents, such as India, Turkey, South Africa, and Malaysia. 

Al Abdullatif Industrial Investment Company is a national Saudi-Arabia-based company specializing in both the distribution and manufacture of weaving products such as blankets, rugs, and carpets, and intermediates such as nylon, polyester, etc. 

The Electrical Industries Company, abbreviated EIC, started in 2005, and since then it has become a leading manufacturer of electrical products to satisfy the growing demand for electrical equipment in Saudi Arabia.

The Thob Al Aseel Company now operates under the Al-Jedaie brand. Since its foundation in 1970, it has been based in the capital Riyadh. It is subscribed under the thobes and fabrics segments and is concerned with wholesale, ready-made clothing and retail fabrics for development, importation, and exportation. It products a variety of products such as underwear, thobes, pajamas, and sleeping robes. It also produces ehrams, T-shirts, and cotton socks. The Thob Al Aseel company is also concerned with selling children's fabrics, adult clothing products, and many other things. 
\subsection{Descriptive statistics of the brands portfolio}
Table \ref{TableDescriptiveStatistics} below shows the descriptive statistics corresponding to prices and sales for the six brands.
\begin{table}[ht]
\caption{Descriptive statistics for brands $B_i$, $1\leq i\leq6$, prices and sales.} \label{TableDescriptiveStatistics}
\hskip-70pt
\begin{tabular}{ccccccccc} 
\hline
Brand&Mean&Median&Min&Max&Std&Skewness&Kurtosis&JB $(h,p)$\\ 
\hline
\multicolumn{8}{c}{The brand prices}\\
$B_1$&161.42&159.2&105.44&225&26.48&0.21&2.37&(1,10$^{-3}$)\\ 
$B_2$&52.91&53.3&36.95&63.7&4.07&$-$0.08&3.15&(0, 0.34)\\
$B_3$&100.25&99.75&67.10&139.2&16.35&0.20&2.52&(1,$10^{-3}$)\\
$B_4$&15.88&12.66&8.15&40.35&8.12&1.76&4.64&(1,$10^{-3}$)\\
$B_5$&20&20.34&14.46&29.4&3.55&0.39&2.35&(1,$10^{-3}$)\\
$B_6$&35.52&31.35&14.88&69.75&16.44&0.65&2.01&(1,$10^{-3}$)\\
\multicolumn{8}{c}{The brands sales}\\
$B_1$ & 153.63&109.31&7.56&4020&258.03&10.70&140.71&(1,$10^{-3}$)\\
$B_2$ &564.28&418.03&34.70&12,140&689.47&9.25&129.83&(1,$10^{-3}$)\\
$B_3$&890.12&512.49&33.30&1.21310&3967.75&27.64&831.52&(1,$10^{-3}$)\\
$B_4$&809.01&223.43&8.24&22,200&1935.15&6.08&51.00&(1,$10^{-3}$)\\
$B_5$&2577.69&1720&111.39&37,490&3161.13&4.89&40.35&(1,$10^{-3}$)\\
$B_6$&243.89&64.04&0.01&5250&570.35&4.70&29.97&(1,$10^{-3}$)\\
\hline
\end{tabular}
\end{table}

The first deductions from Table \ref{TableDescriptiveStatistics} may arise from the Min and Max values, which are widely different for almost all the brand prices and sales, reflecting the existence of aberrant values or anomalies in the market.  In addition, the mean and median values are also different, even widely different, especially for sales. This large range in the prices and sales is more adequately detected and explained by means of time-scale modeling, as apparently it has no logical causes. 

From Table \ref{TableDescriptiveStatistics}, we easily obtain a non-zero skewness for all the brand prices and sales, which leads to rejection of the symmetry hypothesis. All the brand prices and sales are spread to the right of their mean values, except in the $B_2$ price distribution, which has a small skewness and a kurtosis close to 3, indicating that the quasi-normal distribution behavior is hidden for this series of prices. A negative skewness indicates a left-spreading tail for the $B_1$ prices. The kurtosis reflects non-normal behavior for all brand prices and sales, with heavier tails than for a normal distribution. 

In addition, for all the variables $B_1$, $B_3$, $B_4$, $B_5$, and $B_6$, the Jarque--Bera test leads to a returned value of $h=1$ and a returned $p$-value of the order of $10^{-3}$ at the $5\%$ significance level, which indicates a rejection of the null hypothesis.  

We also notice a large standard deviation, indicating a sparse distribution for both prices and sales around their mean values. For some brand prices and sales, the data are very sparsely distributed. This fact may be explained by the use of large differences in prices and sales, and a non-uniform view of the future on the part of both consumers and managers in this market. Consumer demand for special types of products under the same brand leads to an increase in both the volume of sales and prices, without taking into account the equilibrium with other products under the same brand title. 
\subsection{Wavelet multifractal processing of the brands portfolio}
To easily show the time-scale variations (fluctuations, increases, decreases) of these variables and to further understand their behavior according to the time scale, we reproduce the prices and sales for the brands $B_i$, $1\leq i\leq6$, graphically. This yields, among other interpretations, an easy-to-read description of these variables according to the time scale. 

\begin{figure}[H]
\includegraphics[scale=0.80]{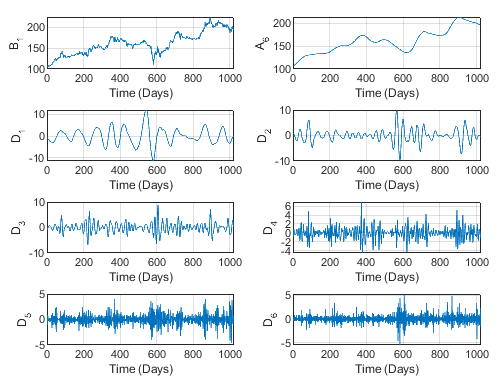}
\caption{The wavelet decomposition of the Jarir brand $B_1$ prices at level 6.}\label{B1PricesWD}
\end{figure}

\begin{figure}[H]
\includegraphics[scale=0.80]{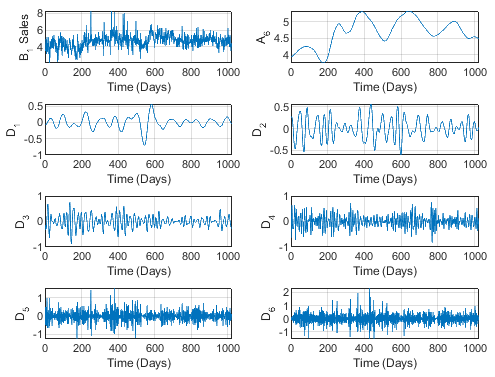}
\caption{The wavelet decomposition of the Jarir brand $B_1$ sales at level 6.}\label{B1SalesWD}
\end{figure}

\begin{figure}[H]
\includegraphics[scale=0.80]{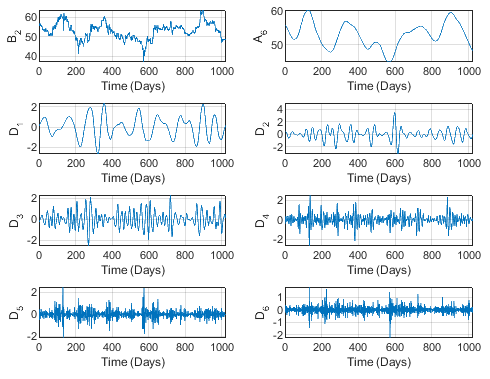}
\caption{The wavelet decomposition of the Almarai brand $B_2$ prices at level 6.}\label{B2PricesWD}
\end{figure}

\begin{figure}[H]
\includegraphics[scale=0.80]{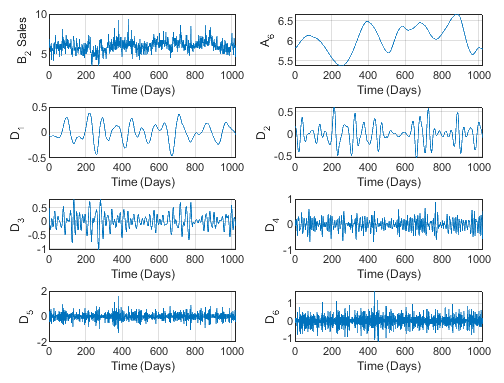}
\caption{The wavelet decomposition of the Almarai brand $B_2$ sales at level 6.}\label{B2SalesWD}
\end{figure}

\begin{figure}[H]
\includegraphics[scale=0.80]{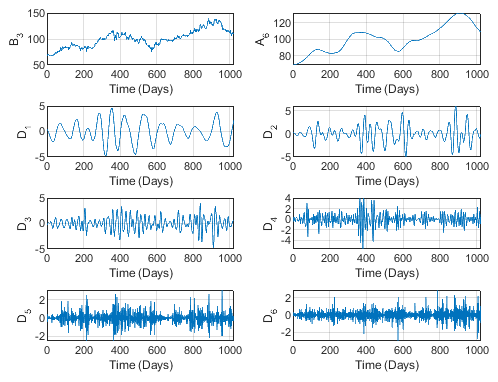}
\caption{The wavelet decomposition of the STC brand $B_3$ prices at level 4.}\label{B3PricesWD}
\end{figure}

\begin{figure}[H]
\includegraphics[scale=0.80]{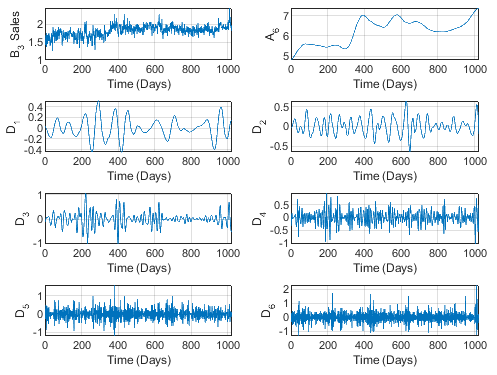}
\caption{The wavelet decomposition of the STC brand $B_3$ sales at level 4.}\label{B3SalesWD}
\end{figure}

\begin{figure}[H]
\includegraphics[scale=0.80]{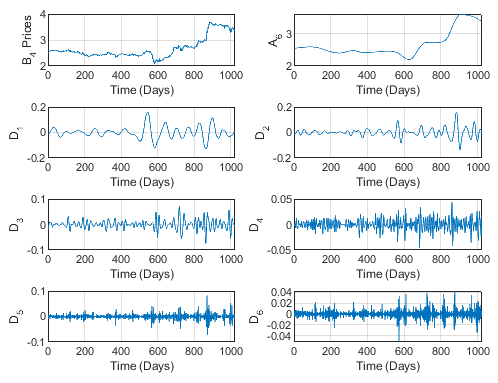}
\caption{The wavelet decomposition of the Al Abdullatif brand $B_4$ prices at level 6.}\label{B4PricesWD}
\end{figure}

\begin{figure}[H]
\includegraphics[scale=0.80]{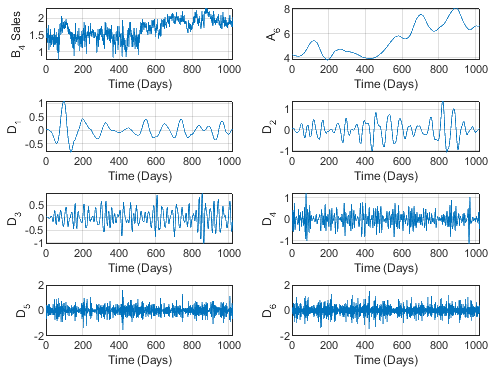}
\caption{The wavelet decomposition of the Al Abdullatif brand $B_4$ sales at level 6.}\label{B4SalesWD}
\end{figure}

\begin{figure}[H]
\includegraphics[scale=0.80]{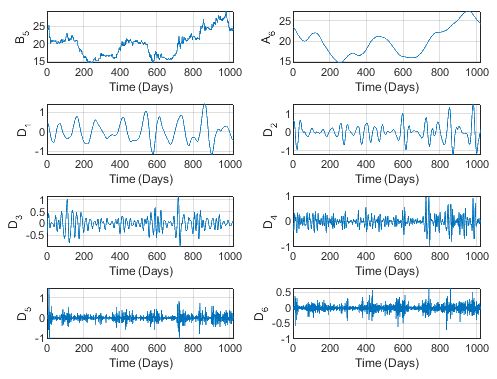}
\caption{The wavelet decomposition of the EIC brand $B_5$ prices at level 6.}\label{B5PricesWD}
\end{figure}

\begin{figure}[H]
\includegraphics[scale=0.80]{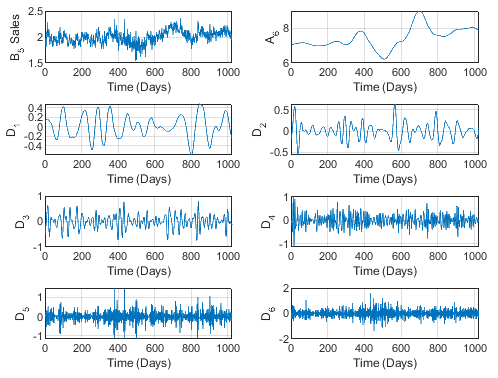}
\caption{The wavelet decomposition of the EIC brand $B_5$ sales at level 6.}\label{B5SalesWD}
\end{figure}

\begin{figure}[H]
\includegraphics[scale=0.80]{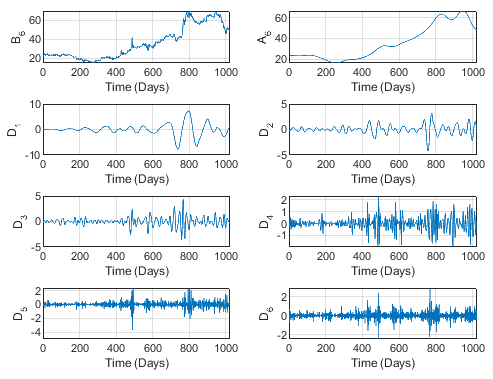}
\caption{The wavelet decomposition of the Al Aseel brand $B_6$ prices at level 6.}\label{B6PricesWD}
\end{figure}

\begin{figure}[H]
\includegraphics[scale=0.80]{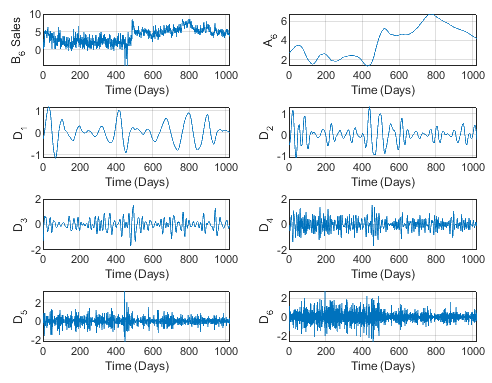}
\caption{The wavelet decomposition of the Al Aseel brand $B_6$ sales at level 6.}\label{B6SalesWD}
\end{figure}

Figures \ref{B1PricesWD}, \ref{B2PricesWD}, \ref{B3PricesWD}, \ref{B4PricesWD}, \ref{B5PricesWD}, and \ref{B6PricesWD} illustrate the wavelet decomposition of the prices for each brand at the level of decomposition $J=6$. In addition, in Figures \ref{B1SalesWD}, \ref{B2SalesWD}, \ref{B3SalesWD}, \ref{B4SalesWD}, \ref{B5SalesWD}, and \ref{B6SalesWD}, we provide the wavelet decomposition due to the brand sales at the same decomposition level of $J=6$. These graphs illustrate the variables with their trends and dynamics or fluctuations. On the $y$-axis, the variable $A_6$ stands for the wavelet approximation at level 6, while the variable $D_i$, $1\leq i\leq6$, stands for the detail component at level $i$. The strong fitting between each variable and its approximation can clearly be seen.
\vspace{-10pt}
\vspace{-10pt}
\vspace{-10pt}
\vspace{-10pt}
\vspace{-10pt}
\vspace{-10pt}
\vspace{-10pt}
\vspace{-10pt}
\vspace{-10pt}
\vspace{-10pt}
\vspace{-10pt}

Figure \ref{B1PricesWD}, illustrating the wavelet decomposition of $B_1$ prices, shows a somewhat increasing behavior, clearly illustrated by the level $6$ trend $A_6$, reminiscent of some perturbations at medium-to-long scales. However, the trend shows no cycles (periodic behavior) for the $B_1$ prices. The perturbation is confirmed in the detail components $D_1$ to $D_6$. According to $D_1, D_2$ a certain pseudo-periodicity seems to take place, with some differences in the maximum prices. This may be explained by the fact that at short time scales (such as weeks in the classical treatments of the time factor), prices are influenced by competitors on the one hand and the global behavior of the market (national as well as international) on the other. One essential factor to be considered is the COVID-19 pandemic, which covered a great part of the period of study and which certainly resulted in decreased sales so that, to compensate for the quantities in stock and thus to recover losses, the company had to increase prices, as in many other cases. This fact is easily shown in the $B_1$ sales graphs in Figure \ref{B1SalesWD}, where we clearly observe an upward trend in sales whenever prices decreased and a downward trend in sales whenever prices increased. In addition, both prices and sales are volatile at long time scales, according to the detail components $D_3$ to $D_6$. We may thus conclude for this brand that short time scales are more comprehensive and easy to understand for consumers, investors, and managers.

In Figure \ref{B2PricesWD}, there is a slight variation around the mean (median), with no exact periodicity but instead a slight pseudo-periodic structure. This behavior is repeated in the sales illustrated in Figure \ref{B2SalesWD}. However, some anomalies appear, essentially at all time scales. This behavior can be naturally understood from the fact that Almarai is a completely national company, based on completely national products. Therefore, compared to $B_1$, for example, its prices and volumes are not strongly affected by importation. In addition, its products are the oldest, the best in the market (according to the majority of consumers), and are consumed daily. All these factors allow price stability in the market, despite a slight growth in the volume of production. In addition, prices and sales seem to reflect the same behavior at short and long time scales. These fluctuations are clearly illustrated by the detail components $D_1$ to $D_6$. Nevertheless, slight fluctuations remain around the mean for all the time scales.

Figures \ref{B3PricesWD} and \ref{B3SalesWD} illustrate the behavior of the brand $B_3$ prices and sales for level 6 wavelet decomposition. Here also, we notice a global increase in both series, which is clearly shown in the trend $A_6$. This increase is sometimes interspersed with small fluctuations. This brand is one of the major suppliers of communications needs in the entire Kingdom of Saudi Arabia.  During the COVID-19 quarantine period, STC sales grew (seen clearly in $A_6$) due to the increase in remote and/or distance communications in almost all domains. Nevertheless, we notice some pseudo-periodicity at short time scales. 

Figures \ref{B4PricesWD} and \ref{B4SalesWD} show quasi-stability at short and medium time scales, followed by a global increase at higher horizons. This increase may also be due to high consumer demand as an alternative during the quarantine period for COVID-19. The pseudo-periodicity and stability appearing at short and medium horizons effectively breaks down at long time scales for these reasons. Moreover, with the lack of importation of similar products, consumers turned to national brands, contributing to the increase in sales and consequently in prices. On the other hand, a high proportion of these types of sales are imported (as raw or elementary commodities). Therefore, with the economic recession afflicting the market for the long COVID-19 period, merchants had to increase prices to compensate for a part of their losses. In addition, other companies use many of the products of this brand, such as polyester and nylon. In critical periods such as during COVID-19, it is necessary to use national stocks. This induces an increase in both sales and prices. 

The brand $B_5$, illustrated by Figures \ref{B5PricesWD} and \ref{B5SalesWD} for prices and sales, respectively, is characterized by a small skewness and kurtosis and quite a small Std. This indicates some stability in both prices and sales, showing some small fluctuations but no important extreme values. As its trend $A_6$ indicates, prices started to decrease at short time scales, turning to an increase at high horizons, with a slow increase at medium levels. Sales seemed to be stable globally at short horizons, decreasing at medium levels, and increasing at longer time scales. Recall that EIC is a national manufacturer of electric instruments. Such instruments need many raw materials, mainly imported from outside the kingdom. Therefore, considering the COVID-19 situation, and with the orientation towards encouraging national and local industry and production being one of the main goals of the Vision 2030 program, these companies have experienced expansion and development in the volumes of sales, as well as in prices. Nevertheless, the graphical representations of the detail components do not reflect any cycles, as might be expected in classical studies.

Figures \ref{B6PricesWD} and \ref{B6SalesWD} illustrate the time-scale behavior of brand $B_6$ prices and sales, according to the wavelet decomposition at level $J-6$. A global increase in both prices and sales can clearly be seen, especially at longer time scales. By investigating the detail components $D_1$ to $D_6$, we may easily exclude the idea of cycles for this series, especially for low/medium horizons. At longer time scales, the series becomes more and more volatile, as for the preceding brands. The increase at the higher horizons is due to the orientation of consumers to local/national products, as there was no importation from outside. Moreover, this brand specializes mainly in national clothing, representing a style that is not widely known outside the GCC countries, although some companies in China target the GCC market. These outside companies discontinued their exports during the COVID-19 period, which therefore resulted in increased sales of national clothing products. In addition, due to the Vision 2030 program, these local/national brands and the producing companies are forced to implement government directions and guidelines to develop and expand national production. 

Overall, these graphs, especially their detail components, reduce the ideas of cycles and stationarity for these marketing series and show instead a volatile behavior, which increases with the time scale. This indicates that the market concerned is emergent, and economic laws must be applied carefully by investors and managers. These facts may have many reasons. One is the COVID-19 crisis, which has been positively used in some cases, especially regarding telecommunications supplies and national products including foods (Almarai) and clothing (Al Aseel). However, this was a disadvantageous situation for many brands relying on imported raw materials, such as EIC. In addition, there was an effect of labor resettlement, which resulted in a considerable amount of non-highly-qualified labor in many sectors. This fact, although it constitutes a principal goal in the Vision 2030 program, may have reduced~productivity.

\begin{figure}[H]
\centering
\includegraphics[scale=0.65]{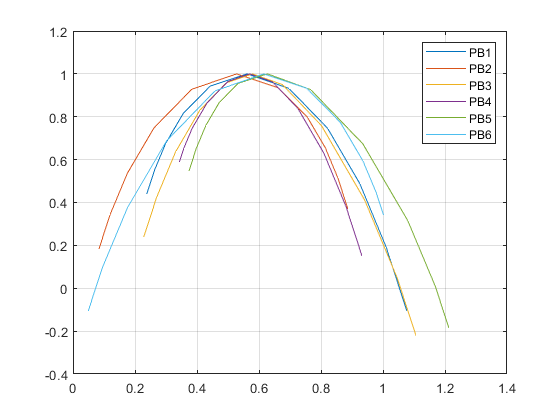}
\caption{The wavelet multifractal spectra of some brands prices.}\label{PBspectra}
\end{figure}

\begin{figure}[H]
\centering
\includegraphics[scale=0.65]{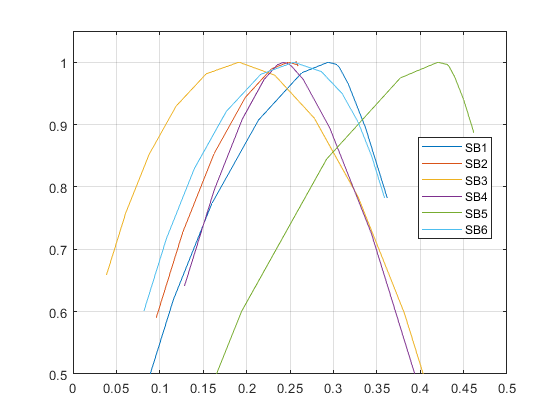}
\caption{The wavelet multifractal spectra of some brands sales.}\label{SBspectra}
\end{figure}

As shown by wavelet analysis of the marketing time series associated to the brands prices and sales, statistical data posses indeed a volatile aspect, where fluctuations become more and more clear over time as shown by the wavelet detail components. To further investigate the multifractal aspect of these series, we subsequently provide the multifractality test based on the multi-fractal wavelet spectrum as shown in Figure \ref{PBspectra} for the brand prices, and Figure \ref{SBspectra} for the sales. We notice easily the high volatile dynamic behavior of the series. 
\vspace{-10pt}
\subsection{The DMD processing of the brands portfolio}
\subsubsection{The time series data processing}
\begin{figure}[H]
\centering
\hspace*{-1cm}
\includegraphics[width=14cm,height=8cm]{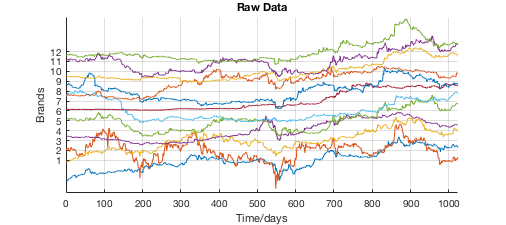}
\caption{The plot displays the raw normalized data of the time series representing the daily price values for the 12 brands. The data spans a period of 1024 days. To enhance visualization, the time series have been vertically stacked along the y-axis.}
\label{rowdataprices}
\end{figure}

\paragraph{Brands prices raw data}
The time series data presented in figure \ref{rowdataprices}, exhibits several interesting patterns. First, there are high-frequency oscillations present in the data, characterized by rapid fluctuations with small amplitudes. These fluctuations suggest short-term market volatility and indicate that prices are subject to frequent and quick changes, potentially influenced by various factors such as news events, market sentiment, or daily trading activities. 

In addition to the high-frequency oscillations, there is a noticeable slow oscillating trend observed in the data. These trend represent longer-term price movements that persist over a significant period. It suggests that the market dynamics underlying these brands exhibit cyclical patterns that extend beyond the day-to-day fluctuations. The presence of such trends can be attributed to factors like seasonal effects, supply and demand imbalances, economic cycles, geopolitical events, or industry-specific conditions that influence the overall market sentiment and investor behavior. Moreover, the long-term behavior of all the brands shows a growing long-term pattern. This observation implies that, on average, the prices of these brands have been increasing over the 1024-day period. This upward trend indicates positive market sentiment and suggests that the market dynamics driving these brands have experienced overall growth and expansion during the observed time frame. 

\begin{figure}[H]
\centering
\hspace*{-1cm}
\includegraphics[width=14cm,height=8cm]{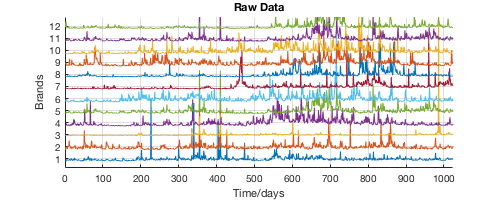}
\caption{The plot displays the raw normalized data of the time series representing the daily sales values for the 12 brands. The data spans a period of 1024 days. To enhance visualization, the time series have been vertically stacked along the y-axis.}
\label{rowdatasales}
\end{figure}

The combination of high-frequency oscillations, slow oscillating trends, and the long-term growing pattern in the price data highlights the complex nature of market dynamics. It signifies the presence of multiple forces at play, including short-term volatility, medium-term cyclical patterns, and long-term growth trends. Understanding and analyzing these patterns can provide valuable insights into the underlying market dynamics, allowing investors and analysts to make informed decisions, manage risk, and identify potential opportunities in the dynamic marketplace.
\paragraph{Brands sales raw data} 
The analysis of the daily sales values for the 12 brands over a period of 1024 days reveals several significant patterns in the data as shown in figure \ref{rowdatasales}. Firstly, the plot of the raw data exhibits high-frequency oscillating patterns with tiny amplitudes for all brands throughout the entire observation period. This suggests that there are rapid fluctuations in sales occurring on a daily basis, but the magnitude of these fluctuations is relatively small for each brand. These high-frequency oscillations may be attributed to factors such as daily consumer behavior, short-term market dynamics, or random noise inherent in the sales data.

Additionally, the plot shows a recurring pattern of spikes that consistently appear across all the time series. However, the height of these spikes varies over different time periods. From day 1 to day 350, the spikes for all brands are relatively low, except for brand 1, which exhibits a notably high spike at day 230. This suggests that there may be a specific event or factor that significantly impacted the sales of brand 1 during that time period, leading to the pronounced spike. From day 350 to day 650, the spikes increase in height for all brands, indicating a period of heightened sales activity or increased demand across the board. From day 650 to day 850, the plot displays quite high spikes for all brands, suggesting a sustained period of strong sales performance or market conditions favoring increased sales. Finally, from day 850 to the last day of the observation period, the height of the spikes decreases again, indicating a decline in sales activity or a shift in market dynamics. Furthermore, the plot reveals a slow oscillating trend with low frequency that persists throughout the entire observation period for all brands. This implies the presence of long-term cyclic behavior or patterns that repeat over extended periods. These slow oscillations could be influenced by factors such as seasonal trends, macroeconomic fluctuations, or industry-wide dynamics that operate on longer time scales. Understanding and analyzing these slow oscillations can provide insights into market dynamics, strategic planning, or identifying brands affected by cyclical trends. The long-term behavior of the plot shows a very slow-growing trend, indicating a gradual increase in sales over time for all brands. This suggests a positive overall market growth or an upward trajectory in consumer demand. However, the rate of growth is relatively slow, indicating a gradual and steady expansion rather than rapid or exponential growth.

In summary, the analysis of the daily sales values for the 12 brands reveals several key patterns. These include high-frequency oscillations with small amplitudes, recurring spikes with varying heights over different time periods, a slow oscillating trend with low frequency, and a very slow-growing long-term trend. These patterns provide insights into the underlying market dynamics, including daily fluctuations, short-term sales variations, event-driven impacts, long-term cyclic behavior, and the overall growth trend. Understanding these dynamics can inform decision-making processes, marketing strategies, and business planning to effectively navigate the market and capitalize on the identified patterns within the sales data.
\subsubsection{The power spectrum}
The DMD power spectrum provides valuable insights into the frequency content and dominant modes of variation in the data. It allows us to analyze the distribution of energy across different frequencies and understand the importance of different oscillatory components in the system. By examining the DMD power spectrum, we can identify the frequencies or modes that contribute the most energy or have the strongest influence on the dynamics of brand prices. This can help us uncover dominant oscillatory patterns, periodic behaviors, or characteristic frequencies in the data.

The power spectrum can reveal the presence of both low-frequency and high-frequency components in the brand price and sales dynamics.  the high-frequency modes provide insights into the short-term dynamics and fast variations of brand prices and sales. These modes capture rapid oscillations or fluctuations in brand prices and sales, which can reveal short-term relationships, interactions, and market dynamics between the brands. By examining the high-frequency modes, one can gain information about the temporal dynamics and synchronization of price/sale changes among the brands at a fast pace. They can reveal whether the brands tend to move together or exhibit divergent behaviors in response to short-term market fluctuations, news events, or other high-frequency factors. Analyzing the high-frequency modes in DMD allows for the identification of fast-changing relationships and market dynamics between brands. It provides insights into the speed at which brands respond to market forces and how they interact with each other in the short run. This information can be valuable for short-term trading strategies, risk management, and understanding market microstructure. Moreover, the high-frequency modes can help identify short-term anomalies, irregularities, or sudden shifts in brand prices and sales. These modes may correspond to transient market inefficiencies, price/sale spikes, or other fast-moving phenomena that can be exploited or monitored for trading opportunities.

On the other hand, the low-frequency modes offer insights into the long-term trends and overall behavior of brand prices/sales. These modes capture the slow variations or trends in brand prices/sales, which can reveal the underlying dynamics and relationships between the brands at longer timescales. By examining the low-frequency modes, one can identify patterns of price/sales changes that occur over extended periods. These trends may indicate common movements or correlations between the prices/sales of different brands. If the low-frequency modes exhibit similar long-term behavior across multiple brands, it suggests a strong correlation or co-movement between their prices/sales at slower timescales.

\begin{figure}[H]
\centering
\hspace*{-1cm}
\includegraphics[width=14cm,height=8cm]{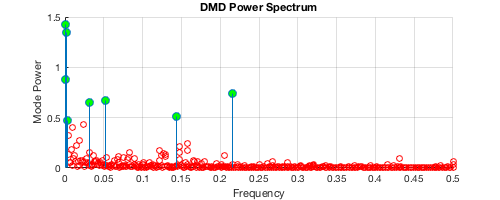}
\caption{The figure shows the DMD spectrum of the time series data presented in figure \ref{rowdataprices}. The DMD spectrum reveals the market-opining price dynamics associated with the brands under consideration.}
\label{Powerprices}
\end{figure}

The low-frequency modes can provide information about the persistent dynamics and synchronization of price/sale changes among the brands. They can reveal whether the brands tend to move together or exhibit divergent behaviors in response to long-term market trends or external factors. Additionally, the low-frequency modes can help identify dominant modes of variation that drive the overall behavior of brand prices/sales. These modes may correspond to fundamental market factors, economic indicators, or other external influences that have a lasting impact on the prices/sales of the brands. Analyzing the low-frequency modes in DMD can be valuable for understanding the macroscopic behavior and relationships between brands in terms of long-term trends, market cycles, and the influence of underlying factors. It provides a broader perspective on the dynamics of brand prices/sales beyond short-term fluctuations, allowing for more strategic decision-making and market analysis.
\paragraph{Brands prices analysis} 
Figure \ref{Powerprices}, shows the DMD power spectrum for the prices time-series data. The modes with the highest power are highlighted in green color. As we will see later in this section, that,  these modes display interesting characteristics and provide valuable insights into the relationship between the brands and the underlying market dynamics. 

The power spectrum analysis provides valuable insights into the underlying driving factors and their implications for market dynamics. Table \ref{tableprices}, shows the frequencies and time duration of completing one cycle, for the selected eight top-ranked modes which have the highest powers.
\begin{table}[ht]
\centering
\caption{The table shows the daily frequencies time duration for completing one full cycle for the selected DMD modes that exhibit the highest powers.}
\begin{tabular}{c|c|c}
\hline
DMD mode & Daily Frequency & Duration of one cycle (days)\\
\hline 
1 & 0.0011 & 925\\
2 & 0.0015 & 654\\
3 & 0.0000 & Inf\\
4 & 0.2154 & 5\\
5 & 0.0517 & 19\\
6 & 0.0320 & 31\\
7 & 0.1436 & 7\\
8 & 0.0030 & 332\\
\hline
\end{tabular}
\label{tableprices}
\end{table}

The presence of high power in very low-frequency modes suggests the existence of long-term trends or cycles that significantly influence the market. These longer-term cycles, such as the 925-day and 654-day cycles, could be driven by factors like macroeconomic conditions, industry-specific events, or changes in investor sentiment that occur over extended periods. Understanding and monitoring these cycles can help investors identify long-term investment opportunities or anticipate major market shifts.

In addition to long-term cycles, the analysis also reveals the importance of shorter-term cycles. The presence of cycles with powers like the 5-day, 19-day, and 31-day cycles indicates the existence of recurring patterns in the market over weekly, fortnightly, and monthly timeframes. These shorter-term cycles may be influenced by factors such as trading patterns, market sentiment, or the release of economic indicators at regular intervals. Traders can leverage these insights to develop short- to medium-term trading strategies, taking advantage of predictable market patterns and timing their trades accordingly.

The identification of a zero-cycle with relatively high power suggests the presence of non-periodic or irregular components impacting market dynamics. These factors could include unexpected events, market shocks, regulatory changes, or other unpredictable influences. Understanding the driving forces behind this non-periodic behavior is crucial for market participants, as it highlights the need to be prepared for sudden shifts in market conditions and the possibility of outliers that can significantly impact investment strategies.

The power spectrum analysis also highlights the relative powers of different cycles, indicating the varying strengths of their influence on market dynamics. Cycles with higher powers, like the 925-day and 654-day cycles, have a more pronounced impact on market behavior, while cycles with lower powers, such as the 7-day and 11-month cycles, have comparatively smaller effects. Recognizing these differences can help investors prioritize their focus and resources, aligning their strategies with the cycles that have the most significant influence on the market.

\begin{figure}[H]
\centering
\hspace*{-1cm}
\includegraphics[width=14cm,height=8cm]{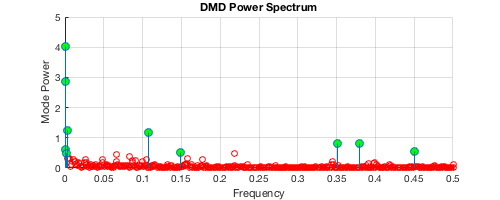}
\caption{The figure shows the DMD spectrum of the time series data presented in figure \ref{rowdatasales}. The DMD spectrum reveals the market-opining sales dynamics associated with the brands under consideration.}
\label{powersales}
\end{figure}

The findings from the power spectrum analysis provide valuable insights into the driving factors and implications for market dynamics. By understanding the recurring patterns at different time scales and the factors driving them, investors can make more informed decisions, anticipate market trends, and adjust their strategies accordingly. It also highlights the importance of considering both long-term and short-term cycles to gain a comprehensive understanding of the market and capitalize on potential opportunities at various time horizons.
\paragraph{Brands sales analysis} 
Figure \ref{powersales}, shows the DMD power spectrum extracted from the time series data presented in figure \ref{rowdatasales}. The modes with the highest power are highlighted in green color. As we will see later in this section, that, these modes display interesting characteristics and provide valuable insights into the relationship between the brands and the underlying market dynamics. Table \ref{tablesales}, shows the frequencies and time duration of completing one cycle, for the selected eight top-ranked modes which have the highest powers.

Upon analyzing the DMD power spectrum for the sales time-series data, several key observations can be made regarding the frequencies associated with the top-ranked modes. The first and second top-ranked modes having zero frequency imply that these modes represent constant or steady components in the sales data. They likely capture the average sales level across all brands without any temporal variation. This could correspond to the baseline sales or the overall market demand that remains relatively stable over time.
\begin{table}[ht]
\centering
\caption{The table shows the daily frequencies and time duration for completing one full cycle for the selected DMD modes that exhibit the highest powers.}
\begin{tabular}{c|c|c}
\hline
DMD mode & Daily Frequency & Duration of one cycle (days)\\
\hline 
1 & 0.0000 &Inf\\
2 & 0.0000 & Inf\\
3 & 0.0038 & 263\\
4 & 0.1081 & 9\\
5 & 0.3511 & 3\\
6 & 0.3801 & 3\\
7 & 0.0010 & 1037\\
8 & 0.4502 & 2\\
9 & 0.1488 & 7\\
10 & 0.0021 & 469\\
\hline
\end{tabular}
\label{tablesales}
\end{table}

The third, seventh, and tenth top-ranked modes having low frequency suggest that they capture sales patterns characterized by long-term trends or slow variations. These modes likely represent market dynamics that evolve over extended periods, such as seasonal trends, economic fluctuations, or changes in consumer behavior. Brands associated with these modes may experience gradual shifts in sales performance, influenced by external factors that operate on longer time scales. Understanding these modes can provide insights into strategic decision-making, long-term forecasting, or identifying brands affected by macro-level market dynamics. On the other hand, the fourth, fifth, sixth, eighth, and ninth top-ranked modes having quite high frequency indicate that they capture sales patterns with rapid fluctuations or short-term variations. These modes are likely driven by factors that operate on shorter time scales, such as weekly, monthly, or quarterly dynamics. Brands associated with these modes may experience more frequent changes in sales performance due to factors like consumer preferences, promotional activities, or periodic events. Understanding these modes can facilitate operational decision-making, inventory management, or identifying brands influenced by more immediate market conditions.

The analysis of the raw data and the findings from the DMD spectrum provide valuable insights into the underlying market dynamics and their connection to the observed patterns. The high-frequency oscillations with small amplitudes identified in the raw data align with the presence of modes in the DMD spectrum that exhibit quite high frequencies. These modes likely capture the rapid fluctuations and short-term variations observed in the sales data, reflecting the daily consumer behavior and immediate market dynamics. Similarly, the recurring spikes in the raw data correspond to certain modes in the DMD spectrum that display distinctive frequencies and magnitudes. The varying heights of these spikes over different time periods in the raw data are indicative of the frequency characteristics associated with specific modes in the DMD spectrum. This connection highlights the ability of the DMD analysis to identify and extract relevant modes that represent the temporal patterns and frequencies inherent in the sales data. Furthermore, the slow oscillating trend with low frequency and the very slow-growing long-term trend observed in the raw data are consistent with the identification of low-frequency modes in the DMD spectrum. These modes capture the long-term cyclic behavior and gradual changes in the sales data, reflecting macroeconomic factors, seasonal trends, or industry-wide dynamics that influence the market over extended periods. Overall, the analysis of the raw data and the insights gained from the DMD spectrum complement each other, providing a comprehensive understanding of the market dynamics, temporal patterns, and frequency characteristics that drive the sales behavior of the brands.
\subsection{Spatial and temporal modes analysis}
\begin{figure}[H]
\centering
\includegraphics[angle=90,scale=0.55]{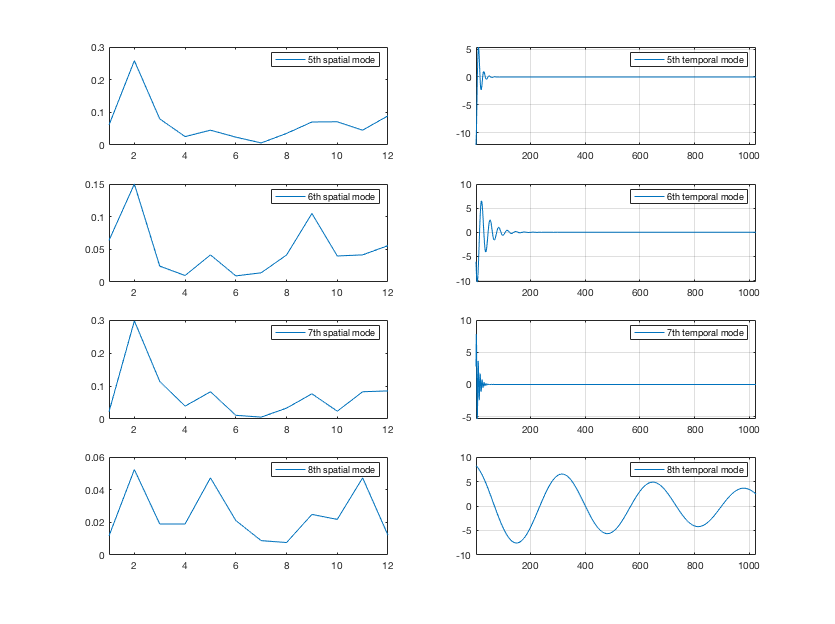}
\caption{The left panel shows the DMD spatial modes, and the right panel shows the corresponding temporal dynamic.}\label{modeprices}
\end{figure}

\begin{figure}[H]
\hspace*{-1.5cm} 
\centering
\includegraphics[width=14.25cm,height=16cm]{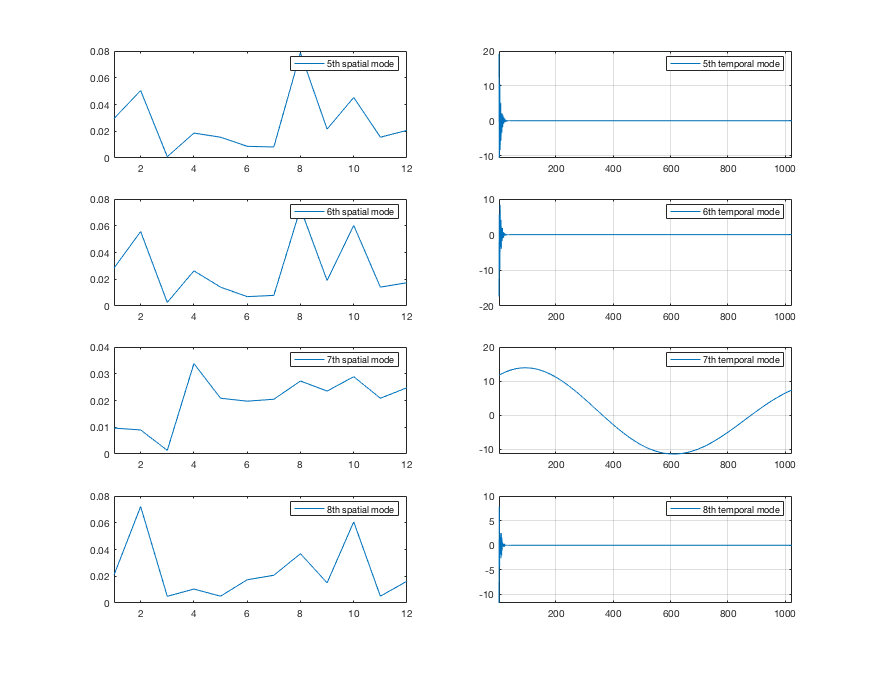}
\caption{The left panels are the DMD spatial modes for the selected top-ranked modes, and in the right panel shows the corresponding temporal dynamic of these modes.}\label{modesales}
\end{figure}

DMD extracts coherent structures or patterns from time-series data and represents them as modes, figure \ref{modeprices} and figure \ref{modesales} show the extracted DMD modes from the brand's prices and sales time series respectively. These extracted structures or patterns may include both short-term and long-term behaviors, capturing the essential dynamics present within the dataset. In the context of brand dynamics, each DMD mode represents a specific pattern or mode of motion that describes how the brands evolve over time. These modes may include synchronized movements, similar trends, or shared dynamics among the brands. By analyzing and interpreting these modes, one can gain insights into the underlying coherent structures or patterns, correlations, and relationships that characterize the dynamics of the brands over time.
\subsubsection{Brands prices analysis}
\paragraph{The first top-ranked mode} 
This mode demonstrates a significant contribution from all brands, with their contributions being relatively comparable, except for brand 2, which exhibits a notably lower alignment with this mode. This observation suggests that the majority of the brands share a common behavior captured by this mode, while brand 2 deviates from this general trend.  Furthermore, the frequency associated with this mode indicates that, this mode exhibits a distinct cyclic behavior with a period of approximately 925 days, which is approximately 2.5 years. This cyclic pattern implies that there is a recurring pattern or process occurring in the market every 925 days, which affects the prices of the brands included in the analysis. 
\begin{figure}[H]
\centering
\includegraphics[angle=90,scale=0.55]{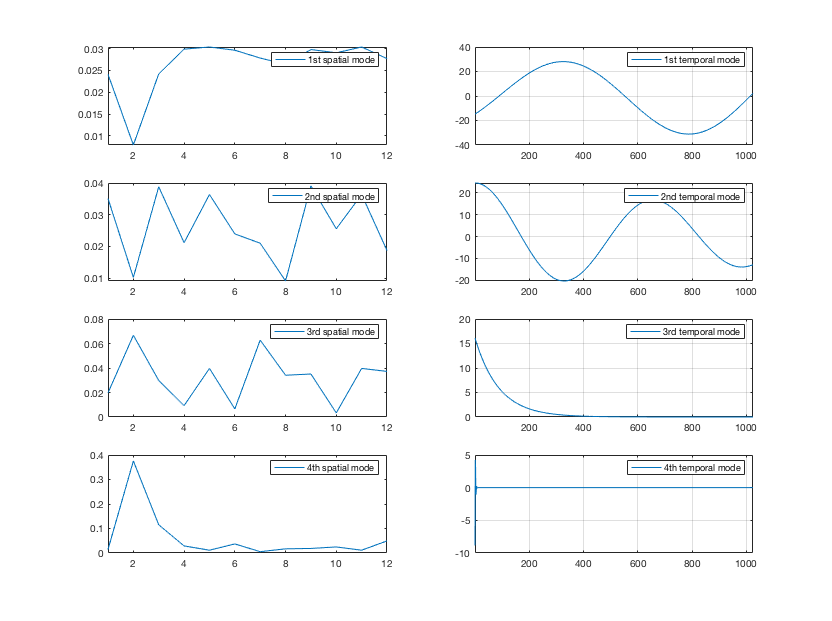}
\caption{The left panel shows the DMD spatial modes and the right panel shows the corresponding temporal dynamic.}\label{cenergyprices1}
\end{figure}

Understanding the nature and drivers of this cycle can provide valuable insights into the longer-term dynamics of the market and may be indicative of specific market events or macroeconomic factors that influence price movements over this timeframe. The associated temporal time dynamic mode, which represents the temporal behavior of this mode, reveals an oscillatory pattern with a relatively high amplitude. Moreover, the frequency of this oscillation is striking, with one complete cycle occurring over the entire 1024-day observation period. This long-term oscillatory behavior suggests the presence of a fundamental market dynamic or underlying factor that influences the prices of the brands. 

The combination of the high contribution from most brands, the cyclic behavior with a 925-day period, and the long-term oscillatory pattern in the temporal mode highlights the interconnectedness of the brands and the underlying market dynamics. The consistent contributions from the majority of the brands suggest a collective response to common market forces or factors that drive their prices. On the other hand, the divergence of brand 2 from this mode may indicate idiosyncratic factors unique to that particular brand or its specific market conditions.
\paragraph{The second top-ranked mode} 
This mode demonstrates a significant contribution from brands 1, 3, 5, 9, 10, and 11, indicating a strong alignment with this mode, and suggesting a collective response to shared market forces or factors that drive their prices in a similar manner. Brands 4, 6, 7, 10, and 12, on the other hand, have a medium contribution to this mode, suggesting a moderate level of alignment, and indicating some commonalities in their price movements. Notably, brand 2 again has the lowest contribution to this mode, indicating a weaker connection with the behavior captured by this mode, and distinct behavior that deviates from the overall market trend captured by this mode. Furthermore, this second mode displays a distinct cyclic behavior with a period of approximately 654 days, which is approximately 1.8 years. This cyclic pattern suggests that there is a recurring pattern or process occurring in the market every 654 days, which influences the prices of the brands included in the analysis. Understanding the nature and drivers of this cycle can provide insights into the medium-term dynamics of the market and may be indicative of specific market events, sectoral trends, or economic indicators affecting the prices of the brands over this timeframe. The associated temporal time dynamic mode, representing the temporal behavior of this mode, exhibits oscillatory behavior with a relatively high amplitude. Moreover, the frequency of this oscillation is noteworthy, with one-and-a-half cycles occurring over the entire 1024-day observation period. This long-term oscillatory behavior suggests the presence of a fundamental market dynamic or underlying factor that influences the prices of the brands captured by this mode. 

The fact that the top-ranked first and second modes have very slow temporal dynamics, indicates a persistent and influential force shaping the market dynamics over the long term. It could reflect factors such as an increase in inflation, also macro-economic factors, such as interest rates, economic growth, fiscal policies, and geopolitical events, which can also influence long-term price trends. 
\paragraph{The third top-ranked mode} 
In this mode brands 2 and 7 exhibit the highest contributions, indicating a strong alignment with this mode, and a significant role played by these brands in shaping the behavior captured by this mode. Brands 1, 3, 5, 8, 9, 11, and 12 contribute to a medium extent, suggesting a moderate level of alignment. On the other hand, brands 1, 4, 6, and 10 have the lowest contributions, indicating a weaker connection with the behavior captured by this mode. Interestingly, this third mode does not exhibit a distinct frequency, where the imaginary part of the associated singular value is zero. The absence of a frequency in this mode suggests that it does not represent any cyclic behavior or recurring pattern within the observed data. Instead, it may capture more subtle or complex dynamics that go beyond periodic trends.

The associated temporal time dynamic mode for this third mode shows a purely moderate decaying behavior. The amplitude of this mode gradually decays over approximately 400 days until it reaches almost zero. This decaying behavior implies that there are underlying factors or forces in the market that gradually dampen the influence or impact of the factors captured by this mode. It suggests that the contribution of the underlying factors driving this mode to the overall market dynamics diminishes over time.
\paragraph{The fourth, fifth, and seventh top-ranked modes} 
In all these modes brand 2 exhibits the highest contribution, indicating a strong alignment with this mode. In contrast, all the other brands have relatively small contributions, suggesting minimal association with the behavior captured by this mode. The fourth mode displays a distinct cyclic behavior with a period of 5 days. This short cyclic pattern implies that there is a recurring pattern or process occurring in the market every 5 days, specifically related to brand 2. Understanding the nature and drivers of this 5-day cycle can provide insights into the intraday dynamics of the market and may be indicative of specific trading patterns, market microstructure effects, or idiosyncratic factors impacting brand 2.  The fifth mode displays a notable cyclic behavior with a period of 19 days. This medium-term cyclic pattern suggests that there is a recurring pattern or process occurring in the market approximately every 19 days, specifically related to brand 2. Understanding the nature and drivers of this 19-day cycle can provide insights into the dynamics of the market over this time frame and may be indicative of specific market events, seasonal trends, or other factors impacting brand 2. The seventh mode displays a distinct cyclic behavior with a period of 7 days. This short-term cyclic pattern suggests that there is a recurring pattern or process occurring in the market approximately every 7 days, specifically related to brand 2. Understanding the nature and drivers of this 7-day cycle can provide insights into the weekly dynamics of the market and may be indicative of specific market events, trading patterns, or other factors impacting brand 2.

The associated temporal time dynamic mode for all these modes exhibits highly decaying oscillatory behavior. The amplitude of the oscillations decays quite rapidly, reaching almost zero over a short period of time. This decaying behavior suggests that the influence of the dynamics captured by these modes diminishes rapidly over time.

Brand 2 plays a significant role in shaping the behavior represented by all these modes, while the other brands have minimal influence. The recurring 5-day, 19-day, and weekly cycles specific to brand 2 suggest that there are intraday, medium-term, and weekly dynamics or trading patterns related to this brand that occur on a regular basis. The rapid decay in amplitude demonstrated by the temporal evolution indicates that these dynamics have a short-term impact and quickly diminish over time. Understanding the dynamics captured by this fifth top mode can provide valuable insights for investors and analysts, particularly those interested in medium-term trading strategies or monitoring brand 2 closely.
\paragraph{The sixth top-ranked mode} 
In this mode, brand 2 exhibits the highest contribution, indicating a strong alignment with this mode. Additionally, Brand 9 has a moderate contribution, while all the other brands have relatively small contributions, suggesting limited association with the behavior captured by this mode. This mode displays a distinct cyclic behavior with a period of 31 days. This medium-term cyclic pattern suggests that there is a recurring pattern or process occurring in the market approximately every 30 days, specifically related to brand 2 and to a lesser extent, brand 9. Understanding the nature and drivers of this 31-day cycle can provide insights into the dynamics of the market over this time frame and may be indicative of specific market events, seasonal trends, or other factors impacting Brand 2 and Brand 9.

The associated temporal time dynamic mode for this mode exhibits decaying oscillatory behavior, transitioning between negative and positive values. The amplitude of this oscillation decays over time, gradually reaching almost zero after several cycles, in about 200 days. This decaying behavior indicates that the influence of the dynamics captured by this mode diminishes over time. The combination of brand 2's highest contribution, Brand 9's moderate contribution, the 31-day cyclic behavior, and the decaying oscillatory pattern in the temporal mode highlight the connection between brand 2, brand 9, and the underlying market dynamics. Brand 2 plays a significant role in shaping the behavior represented by this mode, while Brand 9 also contributes to a moderate extent. The recurring 31-day cycle specific to these brands suggests that there are medium-term dynamics or trends related to these brands that occur periodically. The decay in amplitude indicates that these dynamics have a medium-term impact and gradually diminish over time.

Understanding the dynamics captured by this sixth top mode can provide valuable insights for investors and analysts, particularly those interested in medium-term trading strategies or monitoring brand 2 and brand 9 closely.
\paragraph{The eight top-ranked mode} 
In this mode brands 2, Brand 5, and Brand 11 exhibit the highest contributions, indicating a strong alignment with this mode. Conversely, all the other brands have relatively small contributions, suggesting a limited association with the behavior captured by this mode. This eighth mode displays a distinct cyclic behavior with a period of 11 months. This long-term cyclic pattern suggests that there is a recurring pattern or process occurring in the market approximately every 11 months, specifically related to brand 2, brand 5, and brand 11. Understanding the nature and drivers of this cycle can provide insights into the annual dynamics of the market and may be indicative of seasonal trends, macroeconomic factors, or other long-term influences impacting these brands. The associated temporal time dynamic for this mode exhibits long-term decaying oscillatory behavior, transitioning between negative and positive values. The oscillations complete three cycles in 1024 days, corresponding to the full period of observations. Furthermore, the amplitude of this oscillation is reduced to half its initial value after the full period of observations, indicating a gradual decay of influence over time. This long-term decay in amplitude indicates that these annual dynamic trends have a lasting impact, while gradually diminishing over the full observation period.

This comparative analysis highlights the diverse dynamics and relationships between brands in different modes. It indicates that certain brands consistently contribute more significantly to specific modes, while others have relatively smaller or negligible contributions. Understanding these variations in brand contribution and alignment can provide valuable insights into the underlying market dynamics and help identify key players or factors influencing different modes. Gaining insights from the dominant patterns captured by the highest-ranked modes can be highly valuable for investors and analysts. It allows them to identify prevalent market trends, cyclic patterns spanning short-term, medium-term, and long-term durations, and the underlying factors influencing brand prices during those periods. This knowledge proves particularly useful for those interested in devising investment strategies across different timeframes.

By understanding and recognizing these patterns, market participants can make more informed decisions regarding their short-term, medium-term, and long-term investment strategies. They can effectively manage risks and potentially uncover trading opportunities associated with the observed dynamics. Additionally, studying these modes enables market participants to pinpoint the specific factors or events driving the cyclic behavior exhibited by these prominent modes. This awareness can help them seize short-term, medium-term, and long-term investment opportunities tied to the observed market dynamics.
\subsubsection{Brands sales analysis} 
\paragraph{The first top-ranked mode} 
The high magnitudes of brands 2, 4, and 10 in this mode imply that these specific brands play a significant role in driving the overall sales patterns captured by the mode. Their strong presence suggests that these brands are likely contributing to the majority of sales within the analyzed time series. This could be attributed to various factors such as brand popularity, effective marketing strategies, product differentiation, or consumer preferences favoring these particular brands. The high magnitudes highlight the prominence of these brands in the market and their potential influence on the overall sales dynamics.
\begin{figure}[H]
\centering
\hspace{-1.5cm} 
\includegraphics[width=14.25cm,height=16cm]{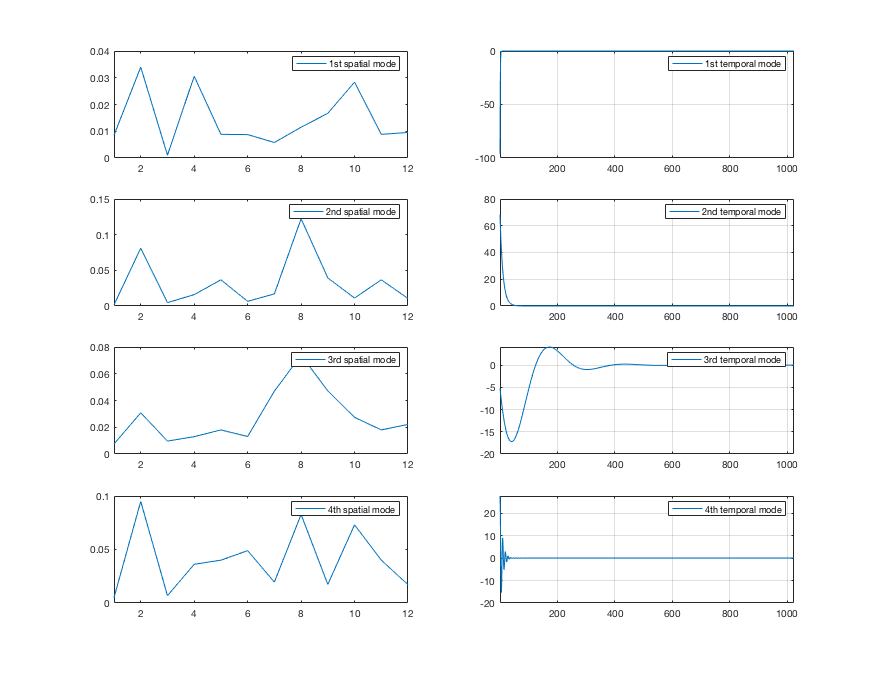}
\caption{The left panels are the DMD spatial modes for the selected top-ranked modes, and in the right panel shows the corresponding temporal dynamic of these modes.}\label{cenergyprices1a}
\end{figure}
\begin{figure}[ht]
\hspace*{-1.5cm} 
\includegraphics[width=14.25cm,height=8cm]{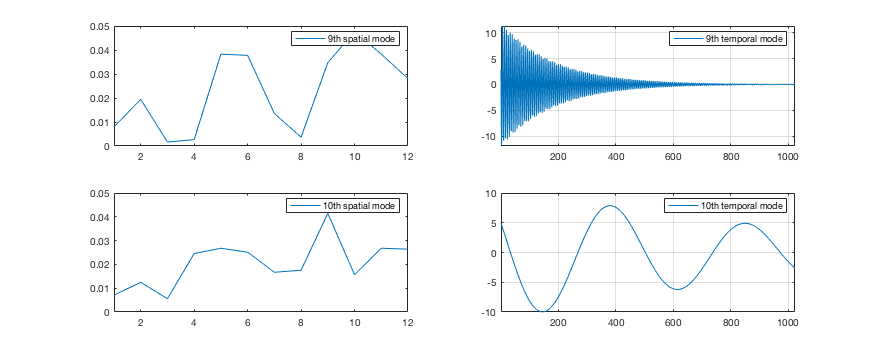}
\caption{The left panel shows the DMD spatial modes, and the right panel shows the corresponding temporal dynamic.}\label{modesalesb}
\end{figure}

On the other hand, the low magnitudes of the remaining brands in this mode indicate that their contributions to the overall sales patterns captured by the mode are relatively minor. These brands might have lower market shares or weaker sales performance compared to brands 2, 4, and 10. The lower magnitudes suggest that the sales of these brands have a relatively smaller impact on the mode's temporal dynamics. Understanding the differences in magnitude among the brands within this mode can provide insights into market segmentation, brand positioning, or competitive dynamics within the industry. The zero frequency of this mode suggests a stationary behavior, indicating that the underlying market dynamics driving the sales patterns in this mode do not exhibit oscillatory or cyclical characteristics. Instead, the mode represents a more stable or consistent aspect of the market dynamics. This could be associated with factors such as long-term brand loyalty, steady demand for specific products, or stable market conditions. The absence of frequency indicates that the sales patterns captured by this mode are driven by non-time-varying factors.

The fast pure decay observed in the temporal dynamic of this mode indicates that the influence or impact of the underlying market dynamics diminishes rapidly over time. This suggests that the factors driving the sales patterns in this mode have a short duration or a limited effect on the brands involved. It could imply that short-term promotional campaigns, temporary market conditions, or specific events are the driving forces behind the sales dynamics captured by this mode. The fast decay highlights the transient nature of the market dynamics associated with this mode, suggesting a need for agile and responsive strategies to capitalize on the identified sales patterns.
\paragraph{The second top-ranked mode} 
The high magnitude of brand 8 in this mode suggests that it plays a significant role in driving the sales patterns captured by the mode. This particular brand likely contributes substantially to the overall sales within the analyzed time series. The high magnitude underscores the prominence of brand 8 in the market and its potential influence on the mode's sales dynamics. Factors such as brand reputation, marketing efforts, product quality, or customer loyalty could contribute to the strong performance of brand 8 within this mode. Understanding the significance of brand 8 provides insights into its market position, competitive advantage, or unique selling proposition. The medium magnitude of brand 2 indicates its moderate contribution to the sales patterns captured by this mode. While not as influential as brand 8, brand 2 still plays a notable role in driving the overall sales dynamics within the mode. This suggests that brand 2 has a relatively stable market presence and a consistent level of consumer demand. Factors such as product portfolio, customer satisfaction, or market positioning could contribute to the moderate performance of brand 2 within this mode. Understanding the medium magnitude of brand 2 allows for insights into its market share, customer base, or competitive positioning. The low magnitudes of the remaining brands in this mode indicate their relatively minor contributions to the overall sales patterns captured by the mode. These brands might have lower market shares or weaker sales performance compared to brand 8 and brand 2. The lower magnitudes suggest that the sales of these brands have a limited impact on the mode's temporal dynamics. Understanding the differences in magnitude among the brands within this mode can provide insights into market segmentation, brand performance, or competitive dynamics within the industry.

Similar to the first top-ranked mode, the zero frequency of the second top-ranked mode indicates a stationary behavior, signifying that the underlying market dynamics driving the sales patterns in this mode do not exhibit oscillatory or cyclical characteristics. Instead, the mode represents a stable or consistent aspect of the market dynamics. This could be associated with factors such as long-term customer preferences, steady demand for specific products, or relatively stable market conditions. The absence of frequency implies that the sales patterns captured by this mode are driven by non-time-varying factors. The fast pure decay observed in the temporal dynamic of this mode suggests that the influence or impact of the underlying market dynamics diminishes rapidly over time. This indicates that the factors driving the sales patterns in this mode have a short duration or a limited effect. It suggests the presence of short-term market trends, transient consumer preferences, or specific events influencing the sales dynamics captured by this mode. The fast decay highlights the need for agile and responsive strategies to capitalize on the identified sales patterns within this mode.
\paragraph{The third top-ranked mode} 
This mode exhibits distinct characteristics that offer insights into the underlying market dynamics and the connection with specific brands. In this mode, the magnitude of brand 8 is high, while the magnitudes of all other brands are relatively low. Furthermore, this mode demonstrates an approximately 9-month cyclic behavior. The associated temporal dynamic of the mode displays a slow decay with a low frequency. The high magnitude of brand 8 in this mode indicates its significant role in driving the sales patterns captured by the mode. Brand 8 likely holds a dominant position in the market, contributing substantially to the overall sales within the analyzed time series. Factors such as brand reputation, customer loyalty, effective marketing strategies, or product differentiation could contribute to the strong performance of brand 8 within this mode. Understanding the prominence of brand 8 provides insights into its market share, competitive advantage, or customer perception. The low magnitudes of the other brands within this mode suggest their relatively minor contributions to the overall sales patterns captured. These brands might have lower market shares or weaker sales performance compared to brand 8. The lower magnitudes indicate that the sales of these brands have a limited impact on the mode's temporal dynamics. Understanding the differences in magnitude among the brands within this mode can provide insights into market segmentation, brand performance, or competitive dynamics within the industry.

The approximately 9-month cyclic behavior observed in this mode implies a recurring pattern in the sales dynamics. This suggests that there are underlying factors or influences that operate on a roughly 9-month cycle, leading to periodic fluctuations in sales. These factors could include seasonal variations, consumer behavior patterns, or industry-specific trends that repeat in a cyclic manner. Understanding the cyclic behavior allows for targeted marketing campaigns, inventory management, or product development strategies to align with the identified sales patterns within this mode.

The slow decay observed in the associated temporal dynamic of this mode indicates that the influence or impact of the underlying market dynamics diminishes gradually over time. This suggests that the factors driving the sales patterns within this mode have a more prolonged effect compared to the previously discussed modes. The slow decay suggests a sustained influence that lingers even after the initial impact. This could be attributed to factors such as long-term consumer preferences, market trends that evolve gradually, or product life cycles. The low frequency of the mode further supports the notion of a slow and sustained influence on the sales dynamics.
\paragraph{The fourth top-ranked mode} 
This mode reveals specific characteristics that provide insights into the underlying market dynamics and the connection with specific brands. In this mode, the magnitudes of brands 2, 8, and 10 are high, while the magnitudes of all other brands are relatively low. Additionally, this mode exhibits an approximately 9-day cyclic behavior. The associated temporal dynamic of the mode displays a fast decay with a high frequency. The high magnitudes of brands 2, 8, and 10 within this mode highlight their significant roles in driving the sales patterns captured. These brands likely hold prominent positions in the market and contribute substantially to the overall sales within the analyzed time series. Factors such as brand reputation, product popularity, effective marketing strategies, or customer loyalty could contribute to the strong performance of brands 2, 8, and 10 within this mode. Understanding the high magnitudes of these brands provides insights into their market shares, competitive advantages, or customer preferences. The low magnitudes of the other brands within this mode suggest their relatively minor contributions to the overall sales patterns captured. These brands might have lower market shares or weaker sales performance compared to brands 2, 8, and 10. The lower magnitudes indicate that the sales of these brands have a limited impact on the mode's temporal dynamics. Understanding the differences in magnitude among the brands within this mode can provide insights into market segmentation, brand performance, or competitive dynamics within the industry.

The approximately 9-day cyclic behavior observed in this mode implies a recurring pattern in the sales dynamics. This suggests that there are underlying factors or influences that operate on a roughly 9-day cycle, leading to periodic fluctuations in sales. These factors could include weekly or bi-weekly consumer behavior patterns, promotional activities, or specific events that impact sales at regular intervals. Understanding the cyclic behavior allows for better timing of marketing campaigns, inventory management, or resource allocation to align with the identified sales patterns within this mode. The fast decay observed in the associated temporal dynamic of this mode indicates that the influence or impact of the underlying market dynamics diminishes rapidly over time. This suggests that the factors driving the sales patterns within this mode have a short duration or a limited effect. The fast decay implies a transient influence that fades quickly after the initial impact. This could be attributed to factors such as short-term promotions, short-lived consumer trends, or time-sensitive market conditions. The high frequency of the mode further supports the notion of a fast-paced and rapidly changing influence on the sales dynamics.
\paragraph{The fifth top-ranked mode} 
This mode exhibits distinct characteristics that provide valuable insights into the underlying market dynamics and the connection with specific brands. In this mode, the magnitude of brand 8 is high, while the magnitudes of brands 1, 2, 9, and 10 are medium. The magnitudes of all other brands within this mode are relatively low. Moreover, this mode demonstrates an approximately 3-day cyclic behavior. The associated temporal dynamic of the mode displays a fast decay with a high frequency. The high magnitude of brand 8 within this mode indicates its significant role in driving the sales patterns captured. Brand 8 likely holds a dominant position in the market, contributing substantially to the overall sales within the analyzed time series. Factors such as brand reputation, customer loyalty, effective marketing strategies, or product differentiation could contribute to the strong performance of brand 8 within this mode. Understanding the prominence of brand 8 provides insights into its market share, competitive advantage, or customer perception. The medium magnitudes of brands 1, 2, 9, and 10 suggest their moderate contributions to the overall sales patterns captured by this mode. These brands likely have relatively stable market positions and consistent levels of consumer demand. Factors such as product quality, customer satisfaction, competitive pricing, or marketing efforts could contribute to the moderate performance of these brands within this mode. Understanding the medium magnitudes of these brands allows for insights into their market shares, customer bases, or competitive positioning. The low magnitudes of the other brands within this mode indicate their relatively minor contributions to the overall sales patterns captured. These brands might have lower market shares or weaker sales performance compared to brand 8 and the aforementioned medium-magnitude brands. The lower magnitudes suggest that the sales of these brands have a limited impact on the mode's temporal dynamics. Understanding the differences in magnitude among the brands within this mode can provide insights into market segmentation, brand performance, or competitive dynamics within the industry.

The approximately 3-day cyclic behavior observed in this mode implies a recurring pattern in the sales dynamics. This suggests that there are underlying factors or influences that operate on a roughly 3-day cycle, leading to periodic fluctuations in sales. These factors could include specific weekly or bi-weekly consumer behavior patterns, short-term promotions, or other temporal patterns that impact sales at regular intervals. Understanding the cyclic behavior allows for targeted marketing campaigns, inventory management, or resource allocation to align with the identified sales patterns within this mode.

The fast decay observed in the associated temporal dynamic of this mode indicates that the influence or impact of the underlying market dynamics diminishes rapidly over time. This suggests that the factors driving the sales patterns within this mode have a short duration or a limited effect. The fast decay implies a transient influence that fades quickly after the initial impact. This could be attributed to factors such as short-term consumer trends, time-sensitive market conditions, or specific events affecting sales within a short timeframe. The high frequency of the mode further supports the notion of a fast-paced and rapidly changing influence on the sales dynamics.
\paragraph{The sixth top-ranked mode} 
This mode reveals distinct characteristics that provide valuable insights into the underlying market dynamics and the connection with specific brands. In this mode, the magnitudes of brand 2, brand 8, and brand 10 are high, while the magnitudes of all other brands are relatively low. Additionally, this mode exhibits an approximately 3-day cyclic behavior. The associated temporal dynamic of the mode displays a fast decay with a high frequency. The high magnitudes of brand 2, brand 8, and brand 10 within this mode highlight their significant roles in driving the sales patterns captured. These brands likely hold prominent positions in the market and contribute substantially to the overall sales within the analyzed time series. Factors such as brand reputation, product popularity, effective marketing strategies, or customer loyalty could contribute to the strong performance of these brands within this mode. Understanding the high magnitudes of these brands provides insights into their market shares, competitive advantages, or customer preferences. The low magnitudes of the other brands within this mode suggest their relatively minor contributions to the overall sales patterns captured. These brands might have lower market shares or weaker sales performance compared to brand 2, brand 8, and brand 10. The lower magnitudes indicate that the sales of these brands have a limited impact on the mode's temporal dynamics. Understanding the differences in magnitude among the brands within this mode can provide insights into market segmentation, brand performance, or competitive dynamics within the industry.

The approximately 3-day cyclic behavior observed in this mode implies a recurring pattern in the sales dynamics. This suggests that there are underlying factors or influences that operate on a roughly 3-day cycle, leading to periodic fluctuations in sales. These factors could include specific weekly or bi-weekly consumer behavior patterns, short-term promotions, or other temporal patterns that impact sales at regular intervals. Understanding the cyclic behavior allows for targeted marketing campaigns, inventory management, or resource allocation to align with the identified sales patterns within this mode.

The fast decay observed in the associated temporal dynamic of this mode indicates that the influence or impact of the underlying market dynamics diminishes rapidly over time. This suggests that the factors driving the sales patterns within this mode have a short duration or a limited effect. The fast decay implies a transient influence that fades quickly after the initial impact. This could be attributed to factors such as short-term consumer trends, time-sensitive market conditions, or specific events affecting sales within a short timeframe. The high frequency of the mode further supports the notion of a fast-paced and rapidly changing influence on the sales dynamics.
\paragraph{The seventh top-ranked mode} 
This mode reveals distinct characteristics that provide valuable insights into the underlying market dynamics and the connection with specific brands. In this mode, the magnitudes of brand 1, brand 2, and brand 3 are low, while the magnitudes of all other brands are high. Additionally, this mode exhibits an approximately 34.5-month cyclic behavior. The associated temporal dynamic of the mode displays a slow decay with a very low frequency, completing only one cycle over the entire observation period of 1024 days. The low magnitudes of brand 1, brand 2, and brand 3 within this mode suggest their relatively minor contributions to the overall sales patterns captured. These brands might have lower market shares or weaker sales performance compared to the other brands within this mode. The low magnitudes indicate that the sales of these brands have a limited impact on the mode's temporal dynamics. Understanding the differences in magnitude among the brands within this mode can provide insights into market segmentation, brand performance, or competitive dynamics within the industry. On the other hand, the high magnitudes of the other brands within this mode highlight their significant roles in driving the sales patterns captured. These brands likely hold prominent positions in the market and contribute substantially to the overall sales within the analyzed time series. Factors such as brand reputation, product popularity, effective marketing strategies, or customer loyalty could contribute to the strong performance of these brands within this mode. Understanding the high magnitudes of these brands provides insights into their market shares, competitive advantages, or customer preferences.

The approximately 34.5-month cyclic behavior observed in this mode implies a long-term recurring pattern in the sales dynamics. This suggests that there are underlying factors or influences that operate on a roughly 34.5-month cycle, leading to periodic fluctuations in sales. These factors could include macroeconomic trends, seasonal variations, industry cycles, or other long-term patterns that impact sales. Understanding the cyclic behavior allows for strategic planning, forecasting, or identifying opportunities and challenges that arise within this specific time frame.

The slow decay observed in the associated temporal dynamic of this mode indicates that the influence or impact of the underlying market dynamics persists over an extended period. This suggests that the factors driving the sales patterns within this mode have a long-lasting effect, potentially spanning multiple years. The slow decay implies a sustained influence that gradually changes over time. This could be attributed to factors such as long-term market trends, changes in consumer preferences, or industry-wide shifts that unfold gradually. The very low frequency of the mode, completing only one cycle over the entire observation period, further emphasizes the long-term nature of the underlying market dynamics.
\paragraph{The eighth top-ranked mode} 
This mode reveals distinct characteristics that provide valuable insights into the underlying market dynamics and the connection with specific brands. In this mode, the magnitudes of brand 2 and brand 10 are high, while the magnitudes of all other brands are relatively low. Additionally, this mode exhibits an approximately 2-day cyclic behavior. The associated temporal dynamic of the mode displays a fast decay with a high frequency. The high magnitudes of brand 2 and brand 10 within this mode highlight their significant roles in driving the sales patterns captured. These brands likely hold prominent positions in the market and contribute substantially to the overall sales within the analyzed time series. Factors such as brand reputation, product popularity, effective marketing strategies, or customer loyalty could contribute to the strong performance of these brands within this mode. Understanding the high magnitudes of these brands provides insights into their market shares, competitive advantages, or customer preferences. The low magnitudes of the other brands within this mode suggest their relatively minor contributions to the overall sales patterns captured. These brands might have lower market shares or weaker sales performance compared to brand 2 and brand 10. The lower magnitudes indicate that the sales of these brands have a limited impact on the mode's temporal dynamics. Understanding the differences in magnitude among the brands within this mode can provide insights into market segmentation, brand performance, or competitive dynamics within the industry.

The approximately 2-day cyclic behavior observed in this mode implies a recurring pattern in the sales dynamics within a short time period. This suggests that there are underlying factors or influences that operate on a roughly 2-day cycle, leading to periodic fluctuations in sales. These factors could include daily consumer behavior patterns, weekday vs. weekend sales variations, or other short-term temporal patterns that impact sales at regular intervals. Understanding the cyclic behavior allows for targeted marketing campaigns, inventory management, or resource allocation to align with the identified sales patterns within this mode.

The fast decay observed in the associated temporal dynamic of this mode indicates that the influence or impact of the underlying market dynamics diminishes rapidly over time. This suggests that the factors driving the sales patterns within this mode have a short duration or a limited effect. The fast decay implies a transient influence that fades quickly after the initial impact. This could be attributed to factors such as daily or short-term consumer trends, time-sensitive market conditions, or specific events affecting sales within a short timeframe. The high frequency of the mode further supports the notion of a fast-paced and rapidly changing influence on the sales dynamics.
\paragraph{The ninth top-ranked mode} 
This mode reveals distinct characteristics that provide valuable insights into the underlying market dynamics and the connection with specific brands. In this mode, the magnitudes of brand 5, brand 6, brand 9, brand 10, brand 11, and brand 12 are high, while the magnitudes of all other brands are relatively low. Additionally, this mode exhibits an approximately 7-day cyclic behavior. The associated temporal dynamic of the mode displays a slow decay with a high frequency. The high magnitudes of brand 5, brand 6, brand 9, brand 10, brand 11, and brand 12 within this mode highlight their significant roles in driving the sales patterns captured. These brands likely hold prominent positions in the market and contribute substantially to the overall sales within the analyzed time series. Factors such as brand reputation, product popularity, effective marketing strategies, or customer loyalty could contribute to the strong performance of these brands within this mode. Understanding the high magnitudes of these brands provides insights into their market shares, competitive advantages, or customer preferences. The low magnitudes of the other brands within this mode suggest their relatively minor contributions to the overall sales patterns captured. These brands might have lower market shares or weaker sales performance compared to brand 5, brand 6, brand 9, brand 10, brand 11, and brand 12.
The lower magnitudes indicate that the sales of these brands have a limited impact on the mode's temporal dynamics. Understanding the differences in magnitude among the brands within this mode can provide insights into market segmentation, brand performance, or competitive dynamics within the industry.

The approximately 7-day cyclic behavior observed in this mode implies a recurring pattern in the sales dynamics with a weekly cycle. This suggests that there are underlying factors or influences that operate on a roughly 7-day cycle, leading to periodic fluctuations in sales. These factors could include weekly consumer behavior patterns, day-of-the-week effects, or other weekly temporal patterns that impact sales at regular intervals. Understanding the cyclic behavior allows for targeted marketing campaigns, inventory management, or resource allocation to align with the identified sales patterns within this mode.

The slow decay observed in the associated temporal dynamic of this mode indicates that the influence or impact of the underlying market dynamics persists over time. This suggests that the factors driving the sales patterns within this mode have a lasting effect beyond the initial impact. The slow decay implies a sustained influence that gradually changes over time. This could be attributed to factors such as weekly consumer habits, long-term trends in demand, or consistent purchasing patterns that unfold over the course of a week. The high frequency of the mode further supports the notion of a weekly cycle and the regularity of the influence on the sales dynamics.
\paragraph{The tenth top-ranked mode} 
This mode reveals distinct characteristics that provide valuable insights into the underlying market dynamics and the connection with specific brands. In this mode, the magnitude of brand 9 is high, while the magnitudes of brands 4, 5, 6, 7, 8, 10, 11, and 12 are medium. On the other hand, the magnitudes of brands 1, 2, and 3 are low. Additionally, this mode exhibits an approximately 16-month cyclic behavior. The associated temporal dynamic of the mode displays a slow decay with quite a low frequency, completing only two cycles over the entire observation period of 1024 days. The high magnitude of brand 9 within this mode highlights its significant role in driving the sales patterns captured. This brand likely holds a prominent position in the market and contributes substantially to the overall sales within the analyzed time series. Factors such as brand reputation, product uniqueness, effective marketing strategies, or customer loyalty could contribute to the strong performance of brand 9 within this mode. Understanding the high magnitude of this brand provides insights into its market share, competitive advantage, or customer preferences. The medium magnitudes of brands 4, 5, 6, 7, 8, 10, 11, and 12 within this mode suggest their relatively moderate contributions to the overall sales patterns captured. These brands might have a sizable market share or a significant customer base but are not as dominant as brand 9. The medium magnitudes indicate that the sales of these brands have a noticeable impact on the mode's temporal dynamics but do not overshadow the influence of brand 9. Understanding the relative magnitudes among these brands within this mode can provide insights into market segmentation, brand performance, or the competitive dynamics within the industry. The low magnitudes of brands 1, 2, and 3 within this mode indicate their relatively minor contributions to the overall sales patterns captured. These brands might have lower market shares or weaker sales performance compared to the other brands within the mode. The low magnitudes suggest that the sales of these brands have limited influence on the mode's temporal dynamics. Understanding the differences in magnitude among these brands within this mode can provide insights into market segmentation, brand performance, or the competitive landscape within the industry.

The approximately 16-month cyclic behavior observed in this mode implies a recurring pattern in the sales dynamics with a long-term cycle. This suggests that there are underlying factors or influences that operate on a roughly 16-month cycle, leading to periodic fluctuations in sales. These factors could include seasonal trends, economic cycles, or other long-term temporal patterns that impact sales at regular intervals. Understanding the cyclic behavior allows for long-term forecasting, strategic planning, or targeted marketing efforts aligned with the identified sales patterns within this mode.

The slow decay observed in the associated temporal dynamic of this mode indicates that the influence or impact of the underlying market dynamics persists over time. This suggests that the factors driving the sales patterns within this mode have a lasting effect beyond the initial impact. The slow decay implies a sustained influence that gradually changes over time. This could be attributed to factors such as long-term consumer trends, macroeconomic conditions, or product life cycles that unfold over the course of several months. The quite low frequency of the mode, completing only two cycles over the whole observation period, further emphasizes the long-term nature of the influence on the sales dynamics.
\subsection{Magnitude and phase analysis}
The elements-wise magnitude and phase-wise information of the DMD mode provide valuable insights into the contributions, alignment, and synchronization of different brands in response to the underlying factors driving the mode. 

The elements-wise magnitude of the DMD mode represents the relative magnitudes of contribution or alignment of the different brands to the mode. Higher magnitudes indicate a stronger influence or contribution of a particular brand to the mode, whereas lower magnitudes suggest a relatively weaker contribution. This information allows us to identify which brands are more closely aligned or have a greater impact on the dynamics captured by the DMD mode. By examining the elements-wise magnitude, we can understand the extent to which each brand's price variations contribute to the overall behavior represented by the DMD mode. Brands with higher magnitudes are more influential in shaping the mode, indicating that their price movements have a greater impact on the observed dynamics. On the other hand, the elements phase-wise of the DMD mode capture the synchronization of the corresponding brands in their response to the underlying factors driving the mode. The phase information represents the relative timing or alignment of the oscillatory patterns among the brands.

Analyzing the elements phase-wise allows us to understand if the brands exhibit synchronized or coordinated movements in response to the underlying factors. If the phase values of the brands are similar or closely aligned, it suggests that they are synchronized and tend to move together in response to the driving factors. Conversely, if the phase values are more dispersed or varied, it indicates that the brands may exhibit divergent behaviors or have different responses to the underlying factors. The phase information provides insights into the temporal coordination and relationships between the brands captured by the DMD mode. It allows us to identify whether the brands tend to move in unison or display varying degrees of synchronization in their response to the underlying dynamics.

By considering both the elements-wise magnitude and phase-wise information, we can gain a comprehensive understanding of the contributions, alignment, and synchronization of different brands in relation to the underlying factors driving the DMD mode. This knowledge can be valuable for identifying key influencers, understanding interdependencies between brands, and predicting their collective behavior in response to market forces or other driving factors.
\begin{figure}[H]
\centering
\includegraphics[angle=90,scale=0.575]{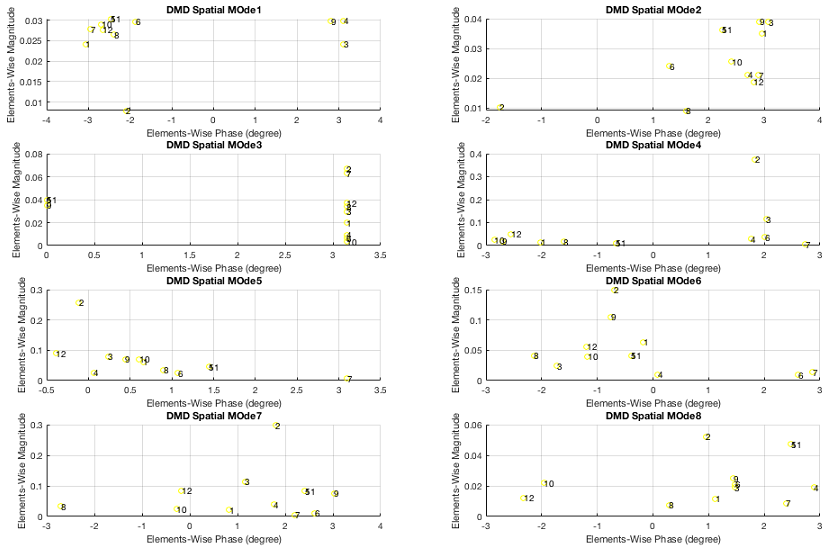}
\caption{The elements-wise magnitude and the corresponding elements-wise phase of the top-ranked brands price DMD modes, are combined and plotted as data points}
\label{magphaseprice}
\end{figure}
\begin{figure}[H]
\centering
\includegraphics[angle=90,scale=0.575]{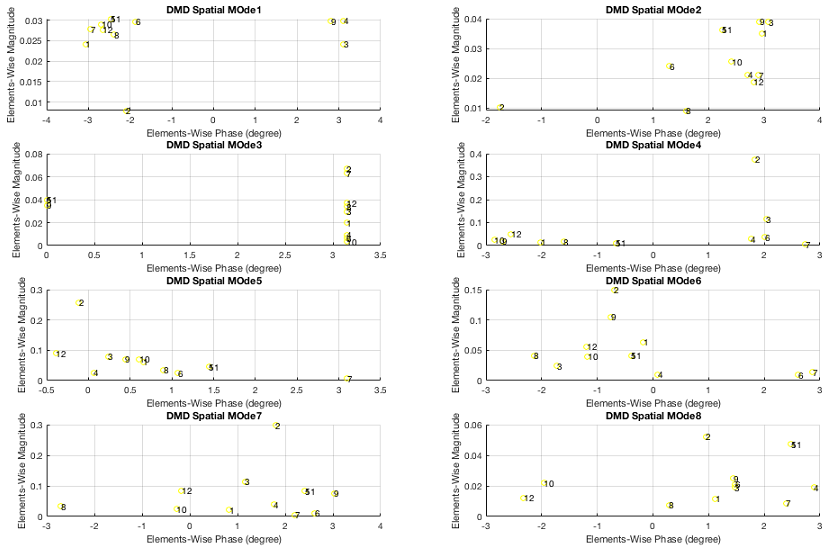}
\caption{The elements-wise magnitude and the corresponding elements-wise phase of the top-ranked brands sales DMD modes, are combined and plotted as data points}
\label{magphasales}
\end{figure}

Figure \ref{magphaseprice} shows the phase-magnitude plot for the eight top-ranked modes, associated with the price time-series data, and figure \ref{magphasales} shows the phase-magnitude plot for the ten top-ranked modes associated with the sales time series data. Across these modes, we observe diverse patterns in the phase and magnitudes of the 12 brands. In some modes, we see brands that stand alone in specific regions, indicating their unique market behavior and potential influence on the overall mode. These brands may be driven by distinct market dynamics, industry-specific factors, or company-specific strategies that set them apart from others. On the other hand, we also notice brands that are spread out in certain regions within the phase-magnitude scatter plots. This suggests that these brands have similarities in their market behavior and may be influenced by common underlying market dynamics or driving factors. The alignment of these brands within the modes indicates a collective contribution to the formation of the respective patterns.
\subsubsection{Brands prices analysis} 
The analysis of the highest power top-ranked mode reveals interesting patterns in terms of the element-wise phase and magnitude distribution among the different brands. The mode exhibits a cyclic behavior with a period of 925 days, completing one full cycle over the entire time frame of 1024 observations. Upon examining the phase and magnitude plot, we observe distinct clusters of brands in different regions. Brands 1, 5, 6, 7, 8, 10, 11, and 12 are tightly clustered in the high magnitude low phase region. This indicates that these brands have a strong alignment and contribute significantly to the dynamics captured by the mode. Their high magnitudes suggest that their price variations have a substantial influence on the overall behavior described by the mode. The low phase values indicate that these brands tend to be synchronized and exhibit similar responses to the underlying factors driving the mode. In contrast, brands 3, 4, and 9 are located in the high magnitude high phase region. This suggests that these brands also contribute significantly to the mode, but their responses to the underlying factors are characterized by a phase shift compared to the tightly clustered brands. The high magnitudes of these brands highlight their impact on the mode, while the high phase values indicate that they exhibit a delayed or shifted response compared to the tightly clustered brands.

Brand 2 stands alone in the low magnitude low phase region. Its low magnitude suggests a relatively weaker contribution to the mode compared to the other brands. The low phase value indicates that this brand exhibits a similar phase alignment to the tightly clustered brands, implying a degree of synchronization. However, its lower magnitude suggests that its price variations have a relatively smaller influence on the overall dynamics captured by the mode. The alignment of these brands with the mode and the underlying market dynamics can provide insights into their interconnectedness and response to common driving factors. The tightly clustered brands in the high magnitude low phase region suggest a strong correlation and synchronized behavior, indicating that they are influenced by shared market forces, economic factors, or other common drivers. Their collective response to these factors contributes significantly to the cyclic behavior captured by the mode. To fully understand the underlying market dynamics and driving factors, a deeper analysis of each brand, industry trends, economic conditions, and other relevant factors would be necessary. However, based on the provided phase-magnitude scatter plots, we can infer that the alignment of brands within the modes and their varying phase and magnitude values indicate a complex interplay of factors shaping their market behavior.
\subsubsection{Brands sales analysis} 
The analysis of the first highest power top-ranked mode, which exhibits a zero frequency and a fast pure decay in its associated temporal dynamics, reveals interesting insights into the alignment of brands with this mode and the underlying market dynamics. In terms of phase and magnitude, brand 2 stands out as it is located in the high magnitude and almost zero phase region. This suggests that brand 2 is strongly aligned with the dynamics captured by this mode. Its high magnitude indicates that brand 2 contributes significantly to the overall behavior described by this mode, while the almost zero phase suggests that its sales patterns are in sync with the dominant oscillatory behavior of this mode. On the other hand, brands 3, 5, and 11 are located in the low magnitude and almost zero phase region. This implies that these brands have relatively weaker contributions to the overall behavior described by the mode and their sales patterns exhibit minimal phase shift compared to the oscillations captured by the mode. These brands may have sales patterns that are less influenced by the dominant oscillatory behavior and may be driven by other factors or dynamics.

The remaining brands (1, 4, 6, 7, 8, 9, 10, and 12) are located in the same high phase region but differ in magnitudes. This indicates that these brands exhibit similar phase shifts in their sales patterns corresponding to the dominant oscillatory behavior of this mode. However, the variation in magnitudes suggests that they contribute differently to the overall behavior described by the mode. Brands 4 and 10 have high magnitudes, indicating that they contribute significantly to the overall behavior captured by the mode. On the other hand, brands 1, 6, 7, 8, 9, and 12 have low magnitudes, suggesting relatively weaker contributions. The connection between these findings and the underlying market dynamics suggests that this mode captures a dominant oscillatory pattern that is shared by several brands. The alignment of brand 2 with high magnitude and almost zero phase indicates that it closely follows the oscillatory behavior described by this mode. This suggests that brand 2 may be strongly influenced by the factors driving this dominant oscillatory pattern in the market.

Brands 3, 5, and 11, with low magnitudes and almost zero phase, exhibit sales patterns that are less influenced by the dominant oscillatory behavior captured by this mode. It is possible that these brands operate in niche markets or are driven by different factors that lead to their distinct behavior. The brands located in the same high-phase region but with different magnitudes indicate that they share a similar phase shift in their sales patterns, aligning with the dominant oscillatory behavior. However, the variation in magnitudes suggests differences in the magnitude of influence or contribution to this behavior. This may imply variations in market share, consumer preferences, or competitive positioning among these brands.
\subsection{Further deductions}
By considering the relationships between the raw data patterns, the characteristics of the spatial and temporal modes, and the underlying market dynamics, businesses can gain valuable insights into brands positioning, market segmentation, and the overall market dynamics that drive sales behavior. Also, this analysis helps to identify the presence of steady components, long-term trends, and the impact of external factors on the sales behavior of the brands.
\subsubsection{Brands prices} 
The analysis of the daily price values of the 12 brands over a 1024-day period reveals that all brands exhibit a noticeable long-term slow-growing trend, except for brand 2, which displays a slow decaying trend. This distinction in the trend patterns indicates that brand 2 operates in a different market dynamic compared to the other brands. The first top-ranked mode in the analysis captures this long-term growing trend observed in the data. It is characterized by a cyclic behavior with a period of approximately 925 days, equivalent to approximately 2.5 years. This suggests the presence of a recurring pattern or influence that drives the long-term growth exhibited by the majority of brands. The associated temporal dynamic of this mode showcases a slow oscillation pattern, completing only one cycle over the entire 1024-day observation period. This slow oscillation further emphasizes the long-term nature of the trend captured by this mode, highlighting that the growth or decay observed in brand prices unfolds gradually over time. Interestingly, the first top-ranked mode has a high contribution from all brands, except for brand 2, which has a relatively small contribution. This indicates that the long-term growing trend is a common feature shared by most brands, except for brand 2. The distinct behavior of brand 2 suggests that it operates under different market dynamics or is influenced by unique factors that result in a slow decaying trend instead of the common growth observed in the other brands. The connection between these findings and the underlying market dynamics suggests that the first top-ranked mode captures the long-term growing trend present in the data. The high contributions from all brands, except brand 2, indicate that this trend is a shared characteristic driven by market-wide factors, such as macroeconomic conditions, industry trends, or consumer behaviors. The contrasting behavior of brand 2, with its slow decaying trend and small contribution to the first mode, suggests that it operates under different market dynamics or factors that set it apart from the common long-term growth observed in the other brands. This distinction implies that brand 2 may be influenced by unique factors, such as specific market conditions, competitive positioning, or consumer preferences, which contribute to its slow decay instead of growth.

Upon analyzing the raw data, it is evident that brand 2 exhibits significant and distinct high-frequency oscillations. These oscillations are not commonly observed in the other brands, indicating that brand 2 experiences unique market dynamics or factors that drive its price fluctuations. These observations are further supported by the fourth, fifth, and seventh top-ranked modes in the analysis. These modes exhibit cyclic behavior with periods of approximately 5, 19, and 7 days, respectively. The presence of these distinct periods suggests the existence of regular patterns or influences on the price fluctuations of brand 2. These cyclic behaviors may be attributed to factors such as seasonal demand, economic indicators, or industry-specific events that impact the pricing dynamics of brand 2. The associated temporal dynamics of these modes exhibit fast decaying oscillation patterns, indicating that the price fluctuations captured by these modes are short-lived. This implies that the market dynamics driving these oscillations may be transient or subject to rapid changes. It suggests that brand 2's pricing behavior is responsive to short-term market forces and may require agile strategies to adapt to the dynamic nature of the market.

Furthermore, these modes show a high contribution from brand 2, suggesting that brand 2 plays a dominant role in driving the cyclic behaviors observed in the price fluctuations. This indicates that brand 2's pricing decisions, market positioning, or consumer demand patterns have a significant influence on the overall market dynamics captured by these modes. It implies that brand 2 holds a unique position in the market, potentially serving as a trendsetter or representing a distinct segment with specific pricing dynamics. The connection between these findings and the underlying market dynamics implies that brand 2 operates within an environment characterized by distinct high-frequency oscillations in its pricing behavior. The cyclic patterns observed in the fourth, fifth, and seventh modes suggest the presence of external factors or market forces that drive these price fluctuations. These dynamics may be influenced by factors such as consumer behavior, market competition, or industry-specific trends.

Also, the analysis of the raw data reveals that brands 2, 8, and 11 exhibit a relatively similar slow oscillating pattern, setting them apart from the other brands. This observation is reflected in the eighth mode, which captures this cyclic behavior characterized by a period of 11 months.
The associated temporal time dynamic of this mode displays a long-term slow decaying oscillatory behavior. Over the 1024-day observation period, the oscillations complete three cycles, indicating a gradual and extended pattern. This suggests that brands 2, 8, and 11 experience similar market dynamics or are influenced by common factors that result in this long-term oscillating behavior. Furthermore, the eighth mode demonstrates a high contribution from brands 2, 8, and 11, while the remaining brands have low contributions. This high contribution from the three brands suggests that they play a significant role in driving the cyclic behavior captured by this mode. Their pricing patterns and market dynamics align closely with the oscillatory nature of this mode. The connection between these findings and the underlying market dynamics implies that brands 2, 8, and 11 share common characteristics or are influenced by similar factors that lead to their apparent similar slow oscillating patterns. These factors could include industry-specific trends, consumer behaviors, or external market forces affecting these brands uniquely.

The high contribution from these brands in the eighth mode suggests that they are key drivers of the cyclic behavior observed. Their pricing decisions, market positioning, or product offerings likely align with the oscillatory nature of this mode, indicating that they are attuned to the market dynamics captured by this mode. In contrast, the low contributions from the other brands indicate that they have different pricing patterns or market dynamics compared to brands 2, 8, and 11. They may be influenced by distinct factors or operate in separate market segments, leading to their divergent behavior from the cyclic patterns captured by the eighth mode.

The analysis of the raw data reveals a noticeable common slow oscillating trend that is prominently observed in brands 1, 3, 5, 9, and 11 while being less evident in brands 4, 6, 7, 10, and 12. Additionally, brands 2 and 8 exhibit distinct slow oscillating patterns, setting them apart from the other brands. These observations are captured by the second top-ranked mode, which characterizes the cyclic behavior with a period of approximately 654 days, equivalent to approximately 1.8 years. This indicates the presence of a recurring pattern or influence that drives the long-term oscillations observed in the prices of the aforementioned brands.

The associated temporal dynamic of this mode showcases a slow oscillation pattern, completing only one and a half cycles over the entire 1024-day observation period. This slow oscillation further emphasizes the long-term nature of the trend captured by this mode, indicating that the growth or decay observed in the prices of the relevant brands unfolds gradually over time. Furthermore, the second top-ranked mode exhibits high contributions from all brands, except for brands 2 and 8, which have negligible contributions. This suggests that the common slow oscillating trend is shared by brands 1, 3, 5, 9, and 11, while brands 2 and 8 exhibit different pricing dynamics that do not align with this mode. The connection between these findings and the underlying market dynamics implies that brands 1, 3, 5, 9, and 11 are subject to common factors or market dynamics that result in the observed slow oscillating patterns. These factors may include industry-specific trends, consumer behaviors, or macroeconomic influences that impact these brands similarly. On the other hand, brands 2 and 8 display distinctive slow oscillating patterns, suggesting that they operate under different market dynamics or are influenced by unique factors that set them apart from the common trend observed in the other brands. Understanding these market dynamics and their connection to specific brands is essential for businesses to make informed decisions. Recognizing the shared slow oscillating trend in brands 1, 3, 5, 9, and 11 allows businesses to identify common market influences and tailor their strategies accordingly. This knowledge can guide pricing decisions, marketing campaigns, and resource allocation to align with the cyclic market dynamics captured by the second mode. Additionally, acknowledging the distinct behaviors of brands 2 and 8 provides insights into their unique market positions and dynamics. Businesses can develop targeted approaches to address the specific factors driving their pricing patterns and market trends, optimizing their strategies to leverage the different market dynamics observed in these brands.
\subsubsection{Brands sales} 
Interestingly, in the analysis of the top-ten ranked spatial modes, a distinct mode emerges with high contributions from almost all brands, except for brands 1, 2, and 3, which have low contributions to this mode, also the temporal dynamics of this mode show a low-frequency pattern. This particular mode is suspected to capture the pattern of the spikes observed in the raw data. The low contributions from brands 1, 2, and 3 to this mode indicate that their sales behavior deviates from the common spike patterns observed in the other brands. This distinction suggests that these three brands may be influenced by unique market dynamics or factors that set them apart from the other brands regarding spike patterns. Analyzing the characteristics of the seventh-highest mode further supports the connection between brands and underlying market dynamics. The connection between these findings and the underlying market dynamics suggests that the spike patterns in the raw data can be attributed to specific factors or events that influence most of the brands, except for brands 1, 2, and 3. Understanding these dynamics and their connection to brands is crucial for market analysis and decision-making. Recognizing the distinct spike patterns and the brands' deviations from common patterns allows businesses to identify unique market factors, consumer preferences, or competitive strategies that impact their sales behavior. 

Besides, in the analysis of the raw data, it becomes evident that brand 3 exhibits a unique pattern that sets it apart from the other brands. This distinct pattern is characterized by a different sales behavior or market dynamics compared to the rest of the brands. This observation is further supported by the finding that brand 3 makes relatively low contributions to the top-ranked spatial modes in the DMD analysis. This suggests that the sales patterns of brand 3 deviate from the common trends captured by these modes, indicating that its dynamics are distinct and less aligned with the overall market behavior. Similarly, brand 1 shows a pattern that is less common among the other brands in the raw data. This implies that brand 1 operates in a different manner or experiences unique market dynamics compared to the other brands. This distinction is reflected in the spatial modes of the top ten ranked modes, as brand 1 exhibits moderate contributions to these modes. This indicates that while brand 1 shares some similarities with the other brands, it also possesses certain unique characteristics or sales patterns that set it apart from the rest of the group. The connection between these observations and the underlying market dynamics suggests that each brand operates within its own market niche or experiences specific factors that drive its sales behavior. Brand 3's distinct pattern and low contributions to the top-ranked spatial modes suggest that it may be influenced by unique market dynamics, targeted consumer segments, or differentiated marketing strategies. Similarly, brand 1's less common pattern and moderate contributions to the spatial modes indicate that it possesses both shared market characteristics and individual brand-specific dynamics. Understanding these brand-specific dynamics is crucial for market analysis and decision-making. Recognizing the unique patterns and contributions of each brand allows for a more comprehensive understanding of the market landscape, enabling businesses to tailor strategies, allocate resources effectively, and make informed decisions that align with the specific dynamics of each brand.
\subsection{Further comments and interpretations}
Economically speaking, the results are of great importance. From a theoretical point of view, they confirm the presence of the time-scale factor or behavior in marketing time series and thus can lead scientists in the field to take this factor into consideration in future models. From a practical point of view, or considering economic and financial reality, the results clearly indicate the movements of the markets studied. We notice stability at short time scales, with a slight growth for many brands. Then, the volumes and the prices become disturbed for medium horizons, sometimes with large fluctuations. Finally, for long horizons (high levels), we notice a remarkable increase. These facts can be justified and explained economically. Indeed, the study period is strongly related to several sociological and political phenomena that had a great influence on the economic state of the market. At the beginning of the period, the local market was affected by the turmoil in several Arab countries, which negatively affected sales, especially since a large part of the workforce originally belonged to these countries. The Qatar embargo and the wars that broke out near the kingdom (forcing it to be involved in some of them) had a great impact on marketing. Thus, the local market lost several tributaries and extensions.

The beginning of the period of study also coincided with the discussion about the added tax on all sales, which negatively affected the purchasing power of citizens and residents and led to a kind of stagnation in the market. Furthermore, some severe laws relating to residents, such as taxes on family members, also led to the permanent deportation of a large number of residents and their families in their countries of origin. This factor had the greatest impact on sales and commercial movements.

As a result of this stagnation, the authorities resorted to ways to compensate for the added value through grants to employees and support for national production, which paralleled the idea of reducing the dependence on oil as a sole income, instead encouraging national industries, of which commercial sales represent a large part. This slowdown contributed to the return of market movements in general and led to an increase in the volume of sales, which in turn led to a recovery in prices. All this coincided with some stability in some Arab Spring countries and the return of normal Gulf relations.

The last factor that should be noted concerns the stability of the local and global situation regarding the COVID-19 pandemic, which has now been brought under control. This led to the reopening of the holy cities (Makkah and Madina) to visitors from all over the world and the return of the labor force to its activities. We recall that the holy cities have the largest human gatherings in the world and are therefore the largest promoters and consumers of national sales, with a significant effect on distribution and publicity for various brands outside the Kingdom of Saudi Arabia.

\section{Conclusion}
In many applications of time series, such as in marketing, classical studies suggest non-stationarity and short-time scales. Nevertheless, studies confirm the existence of a time-frequency aspect. Many mathematical tools and models have been applied by researchers to explain the time-scale behavior of marketing series. Wavelets are the most recent and most powerful tool for this purpose, due to their ability to explain both the time and frequency dependencies in the time series. 

\end{document}